\documentclass[10pt,twocolumn,letterpaper]{article}

\usepackage[pagenumbers]{wacv} 

\usepackage{multirow}
\usepackage{makecell}

\newcommand{\red}[1]{{\color{red}#1}}

\definecolor{wacvblue}{rgb}{0.21,0.49,0.74}
\usepackage[pagebackref,breaklinks,colorlinks,allcolors=wacvblue]{hyperref}

\newcommand{{\nick}}{\textcolor{blue}}

\title{Text Slider: Efficient and Plug-and-Play Continuous Concept Control for Image/Video Synthesis via LoRA Adapters}

\author{Pin-Yen Chiu \quad I-Sheng Fang\quad Jun-Cheng Chen\\
Research Center for Information Technology Innovation, Academia Sinica \\
{\tt\small \{nickchiu, ishengfang, pullpull\}@citi.sinica.edu.tw}
}

\begin{document}

\twocolumn[{
\maketitle
\begin{center}
    \centering
    \captionsetup{type=figure}
    \includegraphics[width=\linewidth]{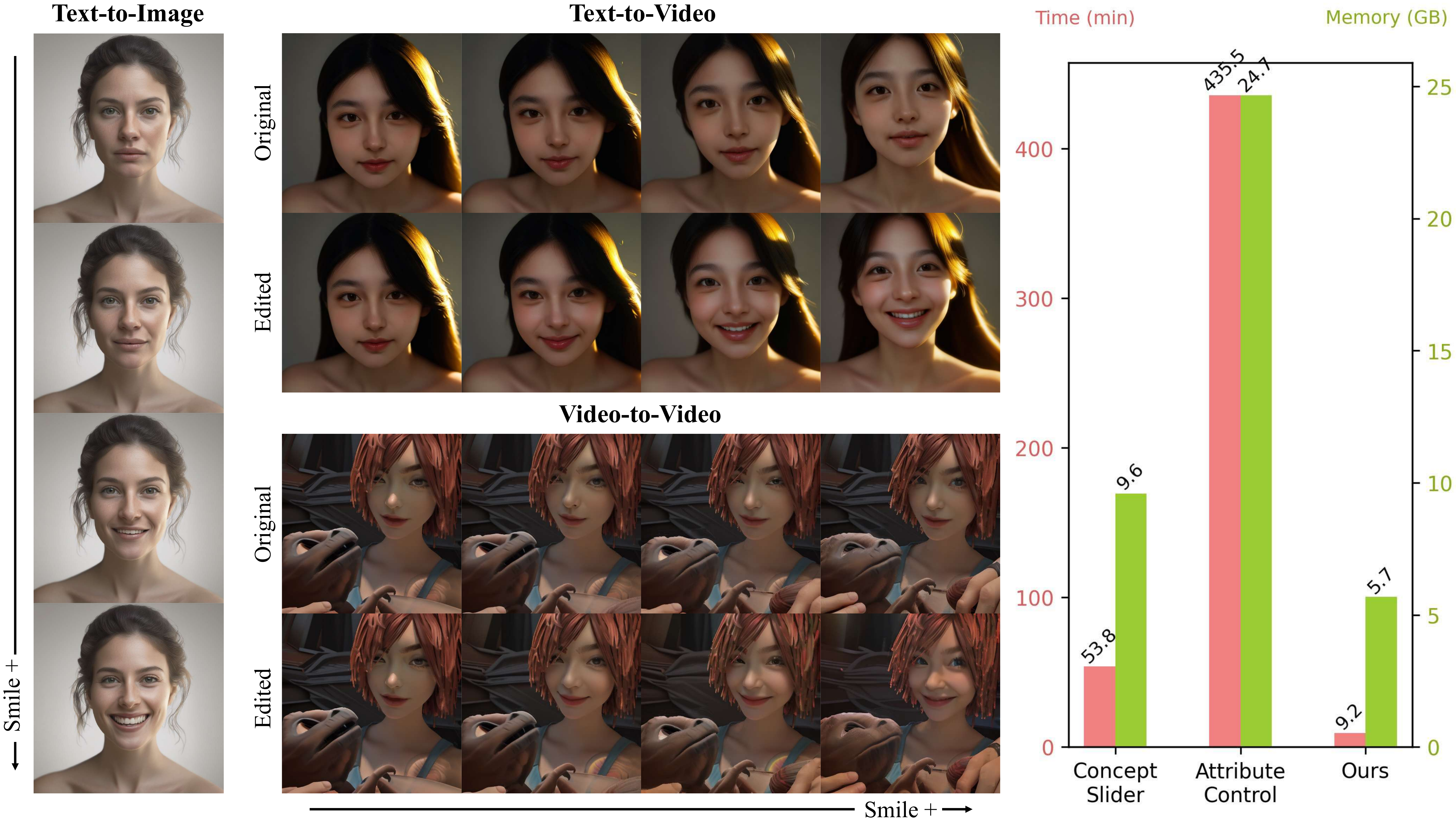}
    \captionof{figure}{Text Slider generalizes effectively to Text-to-Image (SD-XL~\cite{podell2023sdxl}), Text-to-Video (AnimateDiff~\cite{guo2023animatediff}), and Video-to-Video (MeDM~\cite{chu2024medm}) models. It demonstrates strong adaptability across diverse image and video synthesis frameworks without requiring retraining. Compared to Concept Slider~\cite{gandikota2024concept} and Attribute Control~\cite{baumann2025attributecontrol}, Text Slider offers continuous concept control while significantly reducing training time and GPU memory consumption during training, underscoring its lightweight and efficient design. In this example, we manipulate the concept of ``smile'', generating outputs that smoothly transition from a neutral expression to a broad smile. For each video, representative frames are sampled to illustrate the gradual progression of attribute intensity over time.}
    \label{fig:teaser}
\end{center}
}]

\begin{abstract}
Recent advances in diffusion models have significantly improved image and video synthesis. 
In addition, several concept control methods have been proposed to enable fine-grained, continuous, and flexible control over free-form text prompts.
However, these methods not only require intensive training time and GPU memory usage to learn the sliders or embeddings but also need to be retrained for different diffusion backbones, limiting their scalability and adaptability. 
To address these limitations, we introduce Text Slider, a lightweight, efficient and plug-and-play framework that identifies low-rank directions within a pre-trained text encoder, enabling continuous control of visual concepts while significantly reducing training time, GPU memory consumption, and the number of trainable parameters.
Furthermore, 
Text Slider supports multi-concept composition and continuous control, enabling fine-grained and flexible manipulation in both image and video synthesis.
We show that Text Slider enables smooth and continuous modulation of specific attributes while preserving the original spatial layout and structure of the input. 
Text Slider achieves significantly better efficiency: 5$\times$ faster training than Concept Slider and 47$\times$ faster than Attribute Control, while reducing GPU memory usage by nearly 2$\times$ and 4$\times$, respectively. 
Project page:~\url{https://textslider.github.io}.
\end{abstract}    
\vspace{-1.5em}
\section{Introduction}

Diffusion models~\cite{dhariwal2021diffusion, rombach2022high, ho2020denoising} have recently achieved remarkable success in text-guided image and video synthesis~\cite{esser2024scaling, saharia2022photorealistic, flux2024, ho2022video, blattmann2023align}. While text prompts provide flexible and intuitive control, they often fall short in enabling continuous and fine-grained manipulation of specific visual concepts, particularly when subtle variations or intensity levels are required. For instance, expressing nuanced changes in a person’s smile through text alone can be inherently ambiguous. This limitation constrains users’ ability to convey precise visual intent, restricting more expressive and controllable content creation.

Existing methods for continuous and fine-grained concept control are either limited in handling subtle facial attributes or require resource-intensive training with poor adaptability across model architectures.
Prompt-to-Prompt~\cite{hertz2022prompt} achieves localized control by modifying cross-attention maps within latent diffusion models (LDMs)~\cite{rombach2022high}. Video-P2P~\cite{liu2024video} extends this strategy to video domain, but its effectiveness remains limited, especially for subtle facial attributes such as age or smile intensity.
Another line of work, Concept Slider~\cite{gandikota2024concept}, enables concept-specific generation by learning Low-Rank Adapters (LoRA)~\cite{hu2022lora} and modulating a scaling factor during inference. 
However, it suffers from poor adaptability and inefficient training. Specifically, each slider must be separately trained for each diffusion model architecture. For instance, sliders trained for Stable Diffusion 1.5 (SD-1.5)~\cite{rombach2022high} are incompatible with those for Stable Diffusion XL (SD-XL)~\cite{podell2023sdxl} or FLUX.1~\cite{flux2024}. 
These issues significantly hinder its practicality across diverse model architectures, tasks, and large concept sets.
Attribute Control~\cite{baumann2025attributecontrol} proposes another approach by modifying text embeddings to manipulate subject-specific concepts.
However, its optimization-free method has limited performance, while its optimization-based method needs to backpropagate through the diffusion model, resulting in substantial computational costs (\eg, 7 hours and $>$24GB GPU memory for a single concept, see Table~\ref{table:perf-on-t2i}).
Moreover, it struggles with manipulating global attributes such as scene and style, constraining its control flexibility.
To fully unlock the potential of creative and expressive generation, it is essential to develop a more efficient and adaptable method that support continuous attribute modulation, particularly in the context of video synthesis.

In this paper, we introduce Text Slider, a novel approach motivated by prior concept control methods~\cite{gandikota2024concept, baumann2025attributecontrol} and recent advances~\cite{Chen_2024_TEBOpt} that optimize generation by directly tuning token embeddings in the text encoder, without modifying the diffusion model. A key advantage of Text Slider lies in its seamless generalizability across diverse pre-trained diffusion models that share the same text encoder. 
Unlike Concept Slider, which injects low-rank directions into the diffusion model, we achieve similar concept representations more efficiently by fine-tuning LoRA adapters directly within the text encoder, eliminating the need for gradients through the diffusion model.
This design significantly reduces computational requirements, using only $\approx35\%$ of the parameters and $\approx17\%$ of the training time required by Concept Slider on SD-XL, and only $\approx23\%$ of the GPU memory and $\approx2\%$ of the training time compared to Attribute Control on SD-XL, making it accessible for users with consumer-grade GPUs. Furthermore, since the text encoder is shared across multiple diffusion models, Text Slider naturally supports cross-architecture compatibility, enabling continuous concept control across image and video generation.


Our main contributions are summarized as follows:
\begin{itemize}
    \item We propose Text Slider, a method that injects a LoRA module and fine-tune it within the pre-trained text encoder of a diffusion model, enabling continuous concept control without requiring any modification and backpropagation through the diffusion model. Meanwhile, it significantly reduces training time, GPU memory consumption and the number of trainable parameters.
    \item Text Slider is plug-and-play, composable and generalizable across different text-guided image and video diffusion models sharing the same text encoder, making it reusable across architectures.
    \item We extend continuous concept control to the video domain, demonstrating that Text Slider enables continuous and fine-grained attribute manipulation over time while preserving structure and consistency.
\end{itemize}
\section{Related Work}
\subsection{Image and Video Editing}
The use of diffusion models for image and video editing has attracted increasing attention due to their impressive generative capabilities. A common strategy involves providing text prompts to guide edits. However, such approaches often produce entangled modifications, where regions outside the intended target are unintentionally altered. In the image domain, several recent methods~\cite{cao_2023_masactrl, brooks2022instructpix2pix, hertz2022prompt, zhang2023adding, Tumanyan_2023_CVPR} have aimed to improve controllability. For example, InstructPix2Pix~\cite{brooks2022instructpix2pix} fine-tunes the diffusion model to jointly condition on an input image and instruction prompt, and ControlNet~\cite{zhang2023adding} enables fine-grained control by incorporating auxiliary conditions, such as edge maps, depth maps, or keypoints, via an additional trainable branch attached to the diffusion model. In the video domain, early methods~\cite{wu2023tune, liu2024video} typically rely on fine-tuning the model for each video, which is computationally expensive and impractical towards real-time applications. More recent approaches build upon the success of text-guided image diffusion models by extending them to zero-shot video editing~\cite{li2024vidtome, tokenflow2023, chu2024medm, cohen2024slicedit, yang2023rerender}. While these methods can perform object-specific manipulations via text prompts, they rarely explore continuous and fine-grained control over visual attributes in videos. In this work, we address this limitation by introducing a method that enables smooth and precise concept control in video editing. Our approach significantly reduces training overhead and generalizes effectively across diverse model architectures and tasks.

\subsection{Fine-grained and Continuous Concept Control}
Fine-grained and continuous attribute manipulation in generative models has been explored through various paradigms~\cite{kwon2022diffusion, park2023understanding, shen2020interpreting, dalva2024noiseclr, hertz2022prompt, gandikota2024concept, baumann2025attributecontrol, fang2024camera, Lo_2025_ICCV}. One common approach involves discovering semantic directions in latent spaces. Asyrp~\cite{kwon2022diffusion} identifies an intermediate feature space, termed h-space, within the U-Net bottleneck of diffusion models that aligns semantically with CLIP embeddings, allowing attribute control via learned vectors. Similarly, Pullback~\cite{park2023understanding}, inspired by GAN-based editing~\cite{shen2020interpreting}, learns localized basis vectors in the latent space for attribute manipulation. NoiseCLR~\cite{dalva2024noiseclr} adopts an unsupervised contrastive learning approach to extract semantic directions directly within diffusion models. 
Another class of methods leverages attention-based mechanisms. Prompt-to-Prompt~\cite{hertz2022prompt} enables spatially localized edits by modifying cross-attention maps.
Camera Settings as Tokens~\cite{fang2024camera} achieve numerical photographic concept by embedding numerical camera settings into the text embedding space.
More recent works instead rely on contrastive prompt supervision. Concept Slider~\cite{gandikota2024concept} introduces LoRA adapters~\cite{hu2022lora} into the diffusion backbone to modulate attributes via inference-time scaling. While effective, it requires model-specific training and large parameter overhead for each model variant.
Attribute Control~\cite{baumann2025attributecontrol} instead operates in the text embedding space by optimizing a learnable tokenwise embedding using contrastive prompts. Though more parameter-efficient than Concept Slider, it demands backpropagation through the diffusion model and struggles with controlling global attributes like background or style.
In contrast, our method injects LoRA modules into the text encoder and learns concept directions via contrastive text embeddings, eliminating the need to backpropagate through the diffusion model. This significantly reduces training time and GPU memory while enabling efficient, continuous, and flexible control across model architectures and tasks.
\section{Method}

\begin{figure}[t]
    \includegraphics[width=1.0\linewidth]{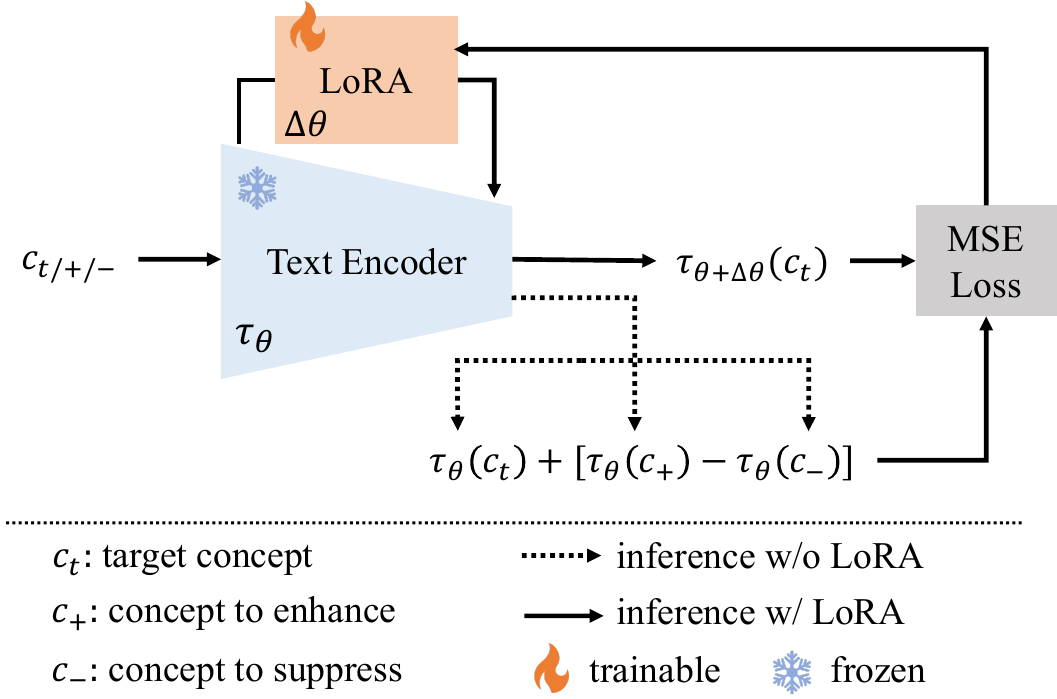}
    \caption{\textbf{Overview of Text Slider.} Text Slider injects and fine-tunes the low-rank parameters $\Delta \theta$ within the pre-trained text encoder $\tau_{\theta}(\cdot)$ of a text-guided diffusion model using contrastive prompts (\eg, $c_{t}$: person, $c_{+}$: old person, and $c_{-}$: young person) derived from concept representations. This enables continuous control over visual attributes across diverse model architectures, supporting both image and video synthesis tasks.}
    \label{fig:system-diagram}
\end{figure}

\subsection{Preliminary}
\noindent\textbf{Text-Guided Diffusion Model.}
Text-to-image generation with diffusion models is achieved by conditioning the generative process on a textual prompt $y$. This prompt is first transformed into a text embedding $\tau_\theta(y)$ using a pre-trained text encoder $\tau_\theta$. The specific text encoder used varies across diffusion models: Stable Diffusion (SD) 1~\cite{rombach2022high} employs the CLIP text encoder~\cite{radford2021learning}; 
SD-XL~\cite{podell2023sdxl} incorporates both CLIP and OpenCLIP.
The resulting text embeddings $\tau_\theta(y)$ serve as conditioning inputs to the diffusion model $\varphi$ via cross-attention layers~\cite{vaswani2017attention}.
Specifically, a cross-attention layer is formulated as $\text{Attn}(Q,K,V) = \text{softmax}\left(\frac{QK^T}{\sqrt{d}}\right) \cdot V$ where the queries $Q$, keys $K$ , and values $V$ are defined as $Q=W_Q \cdot \varphi_i(z_t)$, $K=W_K \cdot \tau_\theta(y)$, and $V=W_V \cdot \tau_\theta(y)$. 
Here, $\varphi_i(z_t)$ denotes intermediate representation of the UNet $\varphi$, and 
$W_Q$,  $W_K$, and $W_V$ are the learnable weight matrices that parameterize the respective linear projections within each cross-attention layer.

\noindent\textbf{Low-Rank Adaptation (LoRA).} 
LoRA~\cite{hu2022lora} is a parameter efficient fine-tuning method that inserts trainable low-rank matrices into pre-trained models while keeping the original weights frozen. Instead of updating the full weight matrix $W_0 \in \mathbb{R}^{d \times k}$, LoRA introduces a low-rank update:
\begin{equation}
    W = W_0 + \alpha \cdot BA,
    \label{eq:lora}
\end{equation}
where $A \in \mathbb{R}^{r \times k}$, $B \in \mathbb{R}^{d \times r}$, and $r \ll \min(d, k)$. The scaling factor $\alpha$ modulates the strength of the update and can be adjusted at inference time to control the influence of the learned direction. \\
\indent In our framework, LoRA is applied to the text encoder, enabling efficient and highly adaptable fine-tuning for continuous concept control in both image and video generation.

\begin{figure*}[t]
    \centering
    \includegraphics[width=1.0\linewidth]{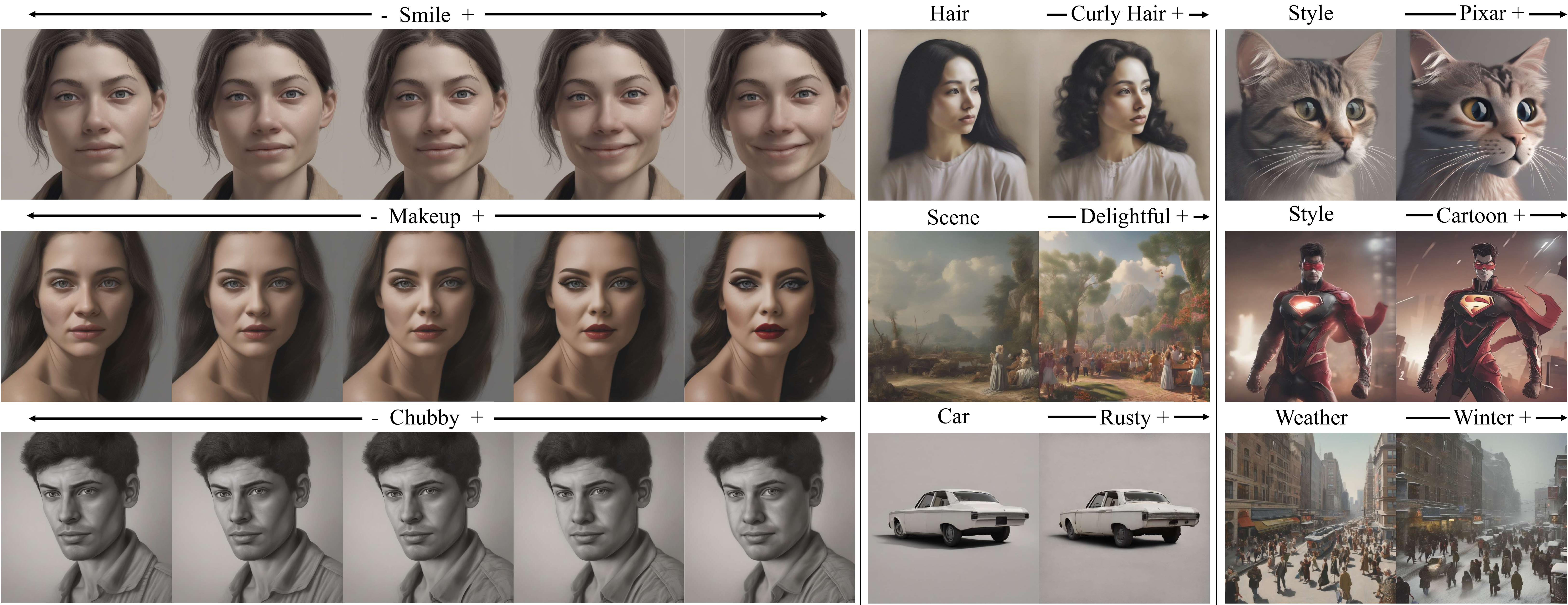}
    \caption{
\textbf{Results on Text-to-Image Generation with SD-XL.} Text Slider enables continuou attribute manipulation across diverse object categories, with controllable attribute intensity achieved by simply adjusting the inference-time scale. Please zoom in for the best view.}
    \label{fig:qualitative-t2i}
\end{figure*}

\begin{table*}
  \resizebox{\linewidth}{!}{
    \begin{tabular}{lcccccccccccc}
        \hline
         & \multicolumn{3}{c}{\textbf{Training}} &\multicolumn{2}{c}{\textbf{Age}} & \multicolumn{2}{c}{\textbf{Smile}} & \multicolumn{2}{c}{\textbf{Curly Hair}} & \multicolumn{2}{c}{\textbf{Chubby}} \\
        \textbf{SD-XL} & Time (s) & Mem. (GB) & \#Params(M) & $\Delta$CLIP ($\uparrow$) & LPIPS ($\downarrow$) & $\Delta$CLIP ($\uparrow$) & LPIPS ($\downarrow$) & $\Delta$CLIP ($\uparrow$) & LPIPS ($\downarrow$) & $\Delta$CLIP ($\uparrow$) & LPIPS ($\downarrow$) \\
        \hline
        Concept Slider~\cite{gandikota2024concept} & 3225.76 & 9.59 & 4.32 & 2.012 & 0.052 & 0.901 & 0.021 & 0.323 & 0.022 & 0.394 & 0.022 \\
        Attribute Control~\cite{baumann2025attributecontrol} & 26132.12 & 24.66 & 0.02 & 2.668 & 0.057 & 2.661 & 0.024 & 0.419 & 0.014 & 0.866 & 0.031 \\
        Text Slider (Ours) & \textbf{550.59} & \textbf{5.68} & 1.53 & 2.904 & 0.072 & 3.832 & 0.035 & 1.285 & 0.022 & 1.292 & 0.029 \\
        \hline
        & \multicolumn{3}{c}{\textbf{Training}} & \multicolumn{2}{c}{\textbf{Age}} & \multicolumn{2}{c}{\textbf{Smile}} & \multicolumn{2}{c}{\textbf{Curly Hair}} & \multicolumn{2}{c}{\textbf{Chubby}}  \\
        \textbf{SD-1.5} & Time (s) & Mem. (GB) & \#Params(M) & $\Delta$CLIP ($\uparrow$) & LPIPS ($\downarrow$) & $\Delta$CLIP ($\uparrow$) & LPIPS ($\downarrow$) & $\Delta$CLIP ($\uparrow$) & LPIPS ($\downarrow$) & $\Delta$CLIP ($\uparrow$) & LPIPS ($\downarrow$) \\
        \hline
        Concept Slider~\cite{gandikota2024concept} & 1263.01 & 3.87 & 2.91 & 2.018 & 0.080 & 2.436 & 0.069 & 2.131 & 0.114 & 0.456 & 0.051 \\
        Attribute Control~\cite{baumann2025attributecontrol} & 26132.12 & 24.66 & 0.02 & 1.470 & 0.035 & 0.918 &0.009 & 0.065 & 0.012& 0.160 & 0.028 \\
        Text Slider (Ours)& \textbf{550.59} & 5.68 & 1.53 & 1.289 & 0.066 & 5.226 & 0.053 & 0.163 & 0.044 & 0.631 & 0.051 \\
        \hline
    \end{tabular}}
  \caption{\textbf{Quantitative Comparison on Text-to-Image Generation.} We evaluate four attributes using a single positive scale for each trained slider. Text Slider achieves competitive performance in both $\Delta$CLIP and LPIPS metrics while substantially reducing training time and GPU memory usage compared to baselines. Notably, it serves as an efficient, plug-and-play solution trained only once and transferable across both SD-XL and SD-1.5 without retraining. For results across multiple scales, we provide $\Delta$CLIP and LPIPS curves over four intensity levels in the Appendix Figure~\ref{fig:main-table-curve-t2i}.}
  \label{table:perf-on-t2i}
  \vspace{-1em}
\end{table*}

\subsection{Text Slider}
Text Slider is a novel method for fine-tuning LoRA adapters on a text encoder~\cite{radford2021learning, cherti2023reproducible} to enable continuous image and video control over designated concepts, as shown in Figure~\ref{fig:system-diagram}. Our approach learns low-rank directions that can enhance or suppress the representation of specific attributes when conditioned on a target concept. 
In contrast to prior works~\cite{gandikota2024concept, baumann2025attributecontrol}, Text Slider does not involve the diffusion model.
By fine-tuning only the LoRA adapters on the text encoder, it generalizes effortlessly across different diffusion architectures and extends naturally to video tasks, without requiring additional training. Furthermore, Text Slider achieves these capabilities with significantly less training time and GPU memory compared to existing methods.

Given a target concept $c_t$, we propose to learn a low-rank direction using a text encoder $\tau_{\theta+\Delta\theta}$ that encourages the expression of more positive attributes $c_+$ while reducing the presence of negative attributes $c_-$. The model $\tau_{\theta+\Delta\theta}$ is trained by minimizing the mean squared error (MSE) between the prompt embeddings from the pre-trained text encoder $\tau_{\theta}$ and the adapted encoder $\tau_{\theta+\Delta\theta}$, using both tokenwise and pooled embeddings. \
Given optimized parameters $\theta^* = \theta+\Delta\theta$, the objective is defined as:
\begin{equation}
    \theta^* = \arg\min_\theta \mathbb{E}_{y} \left\| \tau_t - \tau_{\theta+\Delta\theta}(y) \right\|_2^2.
\end{equation}
As illustrated in Figure \ref{fig:system-diagram}, the target embedding $\tau_t$ is computed as:
\begin{equation}
    \tau_t = \tau_{\theta}(c_t) + \sum_{q \in \mathcal{Q}}(\tau_{\theta}([c_+, q]) - \tau_{\theta}([c_-, q]),
    \label{eq:p_t}
\end{equation}
where $\mathcal{Q}$ is a set of concepts that should be preserved during attribute manipulation. For example, controlling the ``smile'' attribute may unintentionally affect other attributes such as age or gender. By incorporating these preserved concepts into the embedding computation, the learned direction becomes more disentangled and less likely to introduce unwanted changes. 
For more detail information about preserved attributes and the prompts used in training, please refer to our Appendix Table~\ref{table:detail-prompts}.

To achieve varying degrees of editing strength, we utilize a scaling factor $\alpha$ that can be adjusted at inference time within the LoRA formulation (Equation~\ref{eq:lora}). This parameter controls the intensity of the attribute manipulation, allowing for fine-grained edits.

\section{Experiments}

We present the experimental setup in Sec.~\ref{sec:exp_setup}, report the results in Sec.~\ref{sec:exp_results} and Appendix Sec.~\ref{sec:supp-qualitative}, demonstrate generalization ability in Sec.~\ref{sec:generalization}, and provide ablation studies in Sec.~\ref{sec:ablation} and Appendix Sec.~\ref{sec:supp-ablation}. 
We also provide more detailed analysis and experimental results on the counterpart of train-free version of our algorithm for a more comprehensive study in Appendix Sec.~\ref{sec:training-free}.

\subsection{Experimental Setup}
\label{sec:exp_setup}
\begin{figure*}[t]
    \centering
    \includegraphics[width=1.0\linewidth]{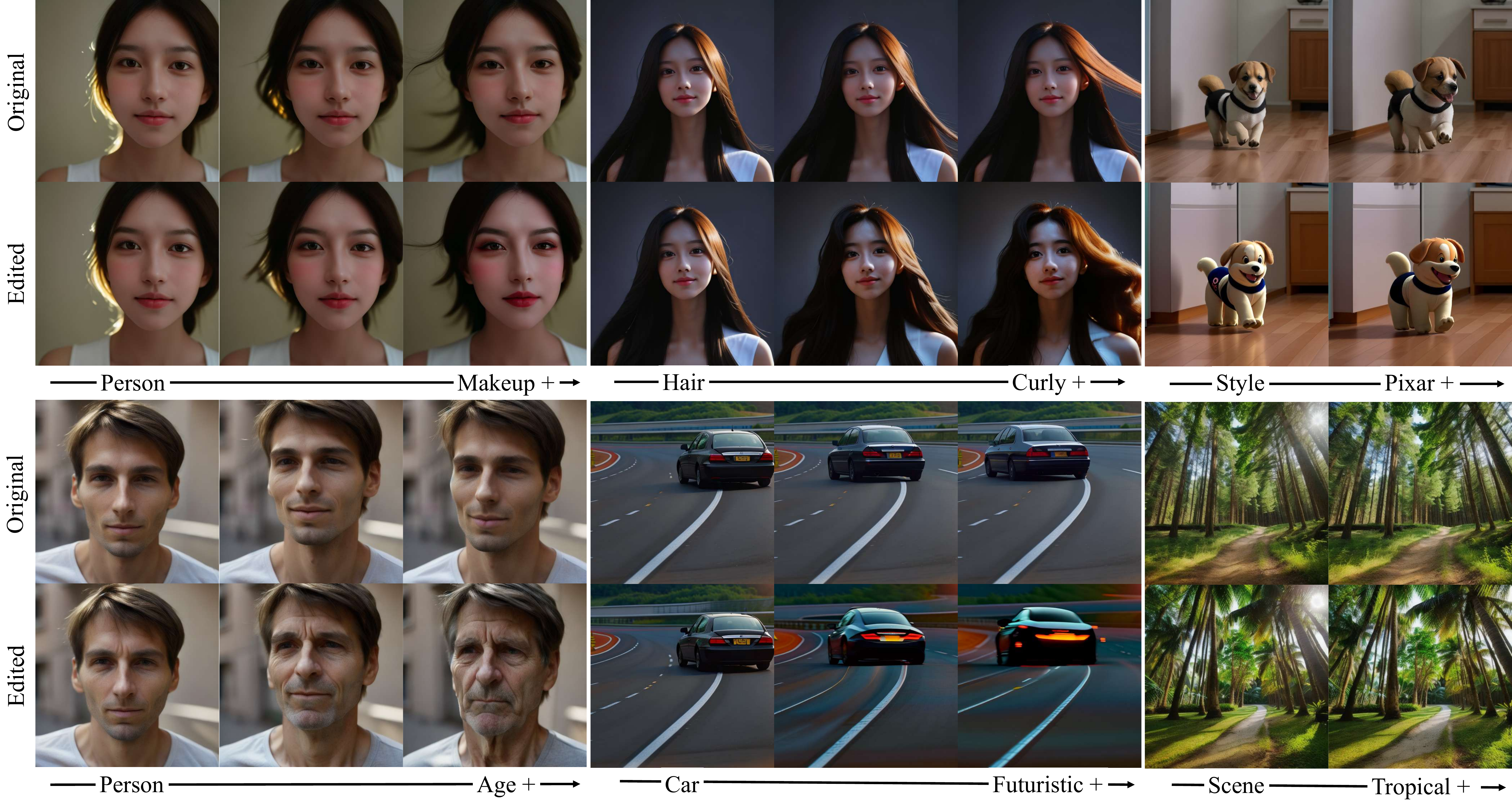}
    \caption{
\textbf{Results on Text-to-Video Generation.} Integrating AnimateDiff~\cite{guo2023animatediff} with Text Slider enables fine-grained and continuous attribute control across diverse object categories, such as person, hair, car, style, and scene, while preserving structural consistency throughout the video. For each video, representative frames are sampled to illustrate the gradual progression of attribute intensity over time.}
    \label{fig:qualitative-t2v}
    \vspace{-1em}
\end{figure*}

\noindent\textbf{Evaluation Metrics.}
We assess the effectiveness of Text Slider using two primary metrics: $\Delta$CLIP~\cite{gandikota2024concept} and LPIPS~\cite{zhang2018perceptual}. $\Delta$CLIP measures attribute control effectiveness by computing the difference in CLIP scores between the original and edited images with respect to a target text prompt that describes the intended edit. LPIPS assesses perceptual similarity between the original and edited images, providing an assessment of visual coherence and content preservation. For each attribute, we generate 1,000 images using a consistent base prompt (e.g., “image of a person, photorealistic”) to ensure fair comparison across methods. In addition to quantitative performance, we also assess efficiency by reporting the training time, GPU memory usage during training, and the number of trainable parameters.

\noindent\textbf{Baselines.}
We compare Text Slider against recent state-of-the-art methods~\cite{baumann2025attributecontrol, gandikota2024concept} for fine-grained concept control in image generation. For video generation, we adopt lightweight frameworks that integrates with these methods across both text-to-video and video-to-video tasks. Specifically, we use AnimateDiff~\cite{guo2023animatediff} as the text-to-video backbone due to its efficiency and compatibility with personalized image diffusion models, and MeDM~\cite{chu2024medm} as the video-to-video backbone, a zero-shot editing framework that applies image diffusion models frame by frame while maintaining temporal consistency.

\noindent\textbf{Implementation Details.}
All experiments are conducted on a single NVIDIA RTX A6000 GPU 48GB. Each Text Slider is trained for 500 epochs using the AdamW optimizer with a learning rate of $2 \times 10^{-4}$ and \texttt{bfloat16} precision. The LoRA rank is fixed at $r = 4$, with adapters applied to the projection layers in all self-attention blocks of the CLIP ViT-L/14 and OpenCLIP ViT/G-14 text encoders. Tokenwise and pooled embeddings from both encoders are concatenated separately, their respective MSE losses are computed, and the results are summed to form the final objective. To preserve the overall structure of the generated results, we adopt the structure-preserving strategy of SDEdit~\cite{meng2022sdedit}. Specifically, LoRA adapters are disabled during the early denoising steps (by setting their multipliers to 0) until timestep $t = 800$, after which they are enabled for the remainder of the denoising process. 

\subsection{Evaluation Results}
\label{sec:exp_results}
We evaluate our method on image synthesis using Stable Diffusion (SD) XL~\cite{podell2023sdxl} and SD-1.5~\cite{rombach2022high}, as well as on text-to-video and video-to-video generation using models built upon SD-1.5 
through qualitative, quantitative and subjective experiments. In addition, we showcase its ability to compose multiple sliders for enhanced control.
Notably, the same Text Slider can be applied across different model architectures and tasks without retraining. For example, by reusing only the CLIP component of Text Slider for fine-grained concept control in SD-1.5.

\begin{figure}[t]
    \centering
    \includegraphics[width=1.0\columnwidth]{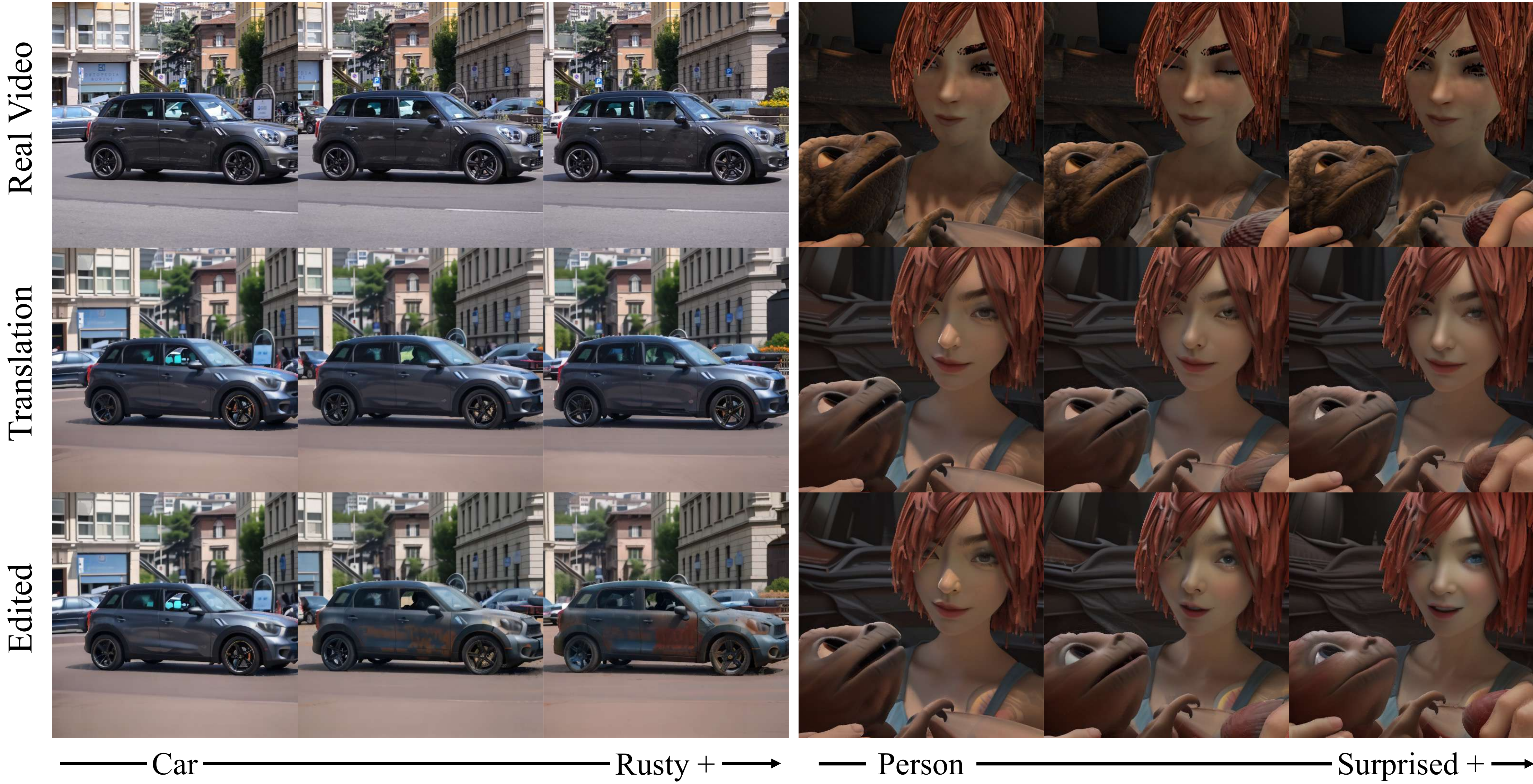}
    \caption{
\textbf{Results on Video-to-Video Generation.} By first translating real videos using MeDM~\cite{chu2024medm} with SDEdit~\cite{meng2022sdedit}, Text Slider enables fine-grained concept control across varying attribute intensities. We demonstrate its effectiveness on different object categories while maintaining structural consistency. For each video, representative frames are sampled to illustrate the gradual progression of attribute intensity over time. Zoom in for the best view.}
    \label{fig:qualitative-v2v}
    \vspace{-1.0em}
\end{figure}

\begin{figure*}[t]
    \centering
    \includegraphics[width=1.0\linewidth]{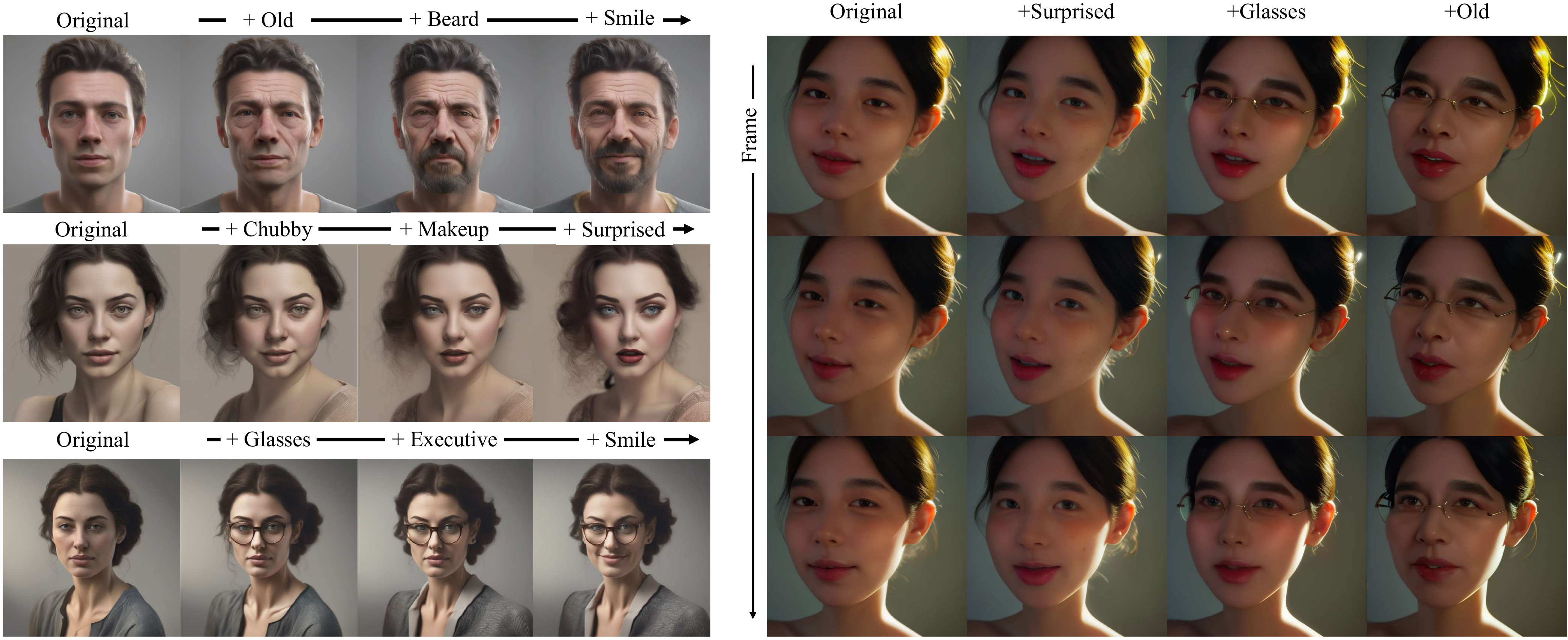}
    \caption{
\textbf{Slider Composition.} We demonstrate the composability of Text Slider in both text-to-image (left) and text-to-video (right) generation by sequentially manipulating different attributes. The proposed approach preserves structural consistency while enabling fine-grained control over the target concepts at each editing stage.}
    \label{fig:compose}
    \vspace{-1em}
\end{figure*}

\noindent\textbf{Text-to-Image Generation.}
We primarily evaluate Text Slider on SD-XL for text-to-image generation. As illustrated in Figure~\ref{fig:qualitative-t2i}, our method enables smooth and continuous manipulation of a wide range of visual attributes across diverse object categories, demonstrating its strong versatility on SD-XL. For quantitative results in Table~\ref{table:perf-on-t2i}, we compare performance on both SD-XL and SD-1.5. Text Slider achieves competitive CLIP scores and LPIPS metrics while substantially reducing training time and GPU memory consumption.
For more diverse results and additional qualitative comparisons with baseline methods, please see Appendix Figure~\ref{fig:more-results}-\ref{fig:more-results-scene} and \ref{fig:t2i-comparison}-\ref{fig:t2i-comparison-sd1} respectively. 

\noindent\textbf{Text-to-Video Generation.}
We qualitatively evaluate the integration of AnimateDiff~\cite{guo2023animatediff} with Text Slider by selecting a diverse set of attributes across object categories including person, hair, car, style, and scene. As illustrated in Figure~\ref{fig:qualitative-t2v}, Text Slider enables continuous modulation of attribute intensity while maintaining spatial coherence. Please see Appendix Figure~\ref{fig:t2v-comparison} for more detailed comparison with baseline methods. 

\noindent\textbf{Video-to-Video Generation.}
To qualitatively assess the effectiveness of combining MeDM~\cite{chu2024medm} with Text Slider, we present results in Figure~\ref{fig:qualitative-v2v} across two attributes: rusty effect on a car and surprised facial expression. Text Slider enables fine-grained attribute modulation while preserving spatial structure and temporal consistency. 
To see the comparison with baseline methods, please see Figure~\ref{fig:v2v-comparison} in Appendix. 

\noindent\textbf{Composing Sliders.}
In Figure~\ref{fig:compose}, we present the qualitative results of composing multiple sliders in text-to-image generation with SD-XL and text-to-video generation using AnimateDiff~\cite{guo2023animatediff}. By sequentially applying various attributes, our method preserves structural consistency at each stage while effectively modulating the intended concepts.

\noindent\textbf{User Study.}
We conduct a subjective evaluation involving 59 participants. Each evaluator is presented with four sets of generated images and videos edited by Text Slider, Concept Slider~\cite{gandikota2024concept}, and Attribute Control~\cite{baumann2025attributecontrol}. Each set includes edits with four attribute intensities across three person-related attributes, smile, age, and chubby, to ensure fair and consistent comparisons. In total, participants answer 12 evaluation questions.
For each question, participants assess results based on three criteria: attribute control effectiveness, smoothness of transition across attribute strengths, and content preservation after editing. Ratings are given on an absolute scale of 1 to 5 (lowest to highest) for each method per criterion. The evaluation spans text-to-image, text-to-video, and video-to-video tasks. The first two sets assess SD-XL and SD-1.5 models respectively for text-to-image. The third set evaluates text-to-video results, including prompt-only AnimateDiff~\cite{guo2023animatediff} baseline. The fourth evaluates video-to-video results, incorporating Video-P2P~\cite{liu2024video} as baseline.
As shown in Table~\ref{table:user-study}, Text Slider outperforms all baselines across all criteria. In addition to superior perceptual quality, our method substantially reduces computational overhead, highlighting its efficiency, adaptability, and practicality.

\begin{figure}[t]
    \centering
    \includegraphics[width=1.0\columnwidth]{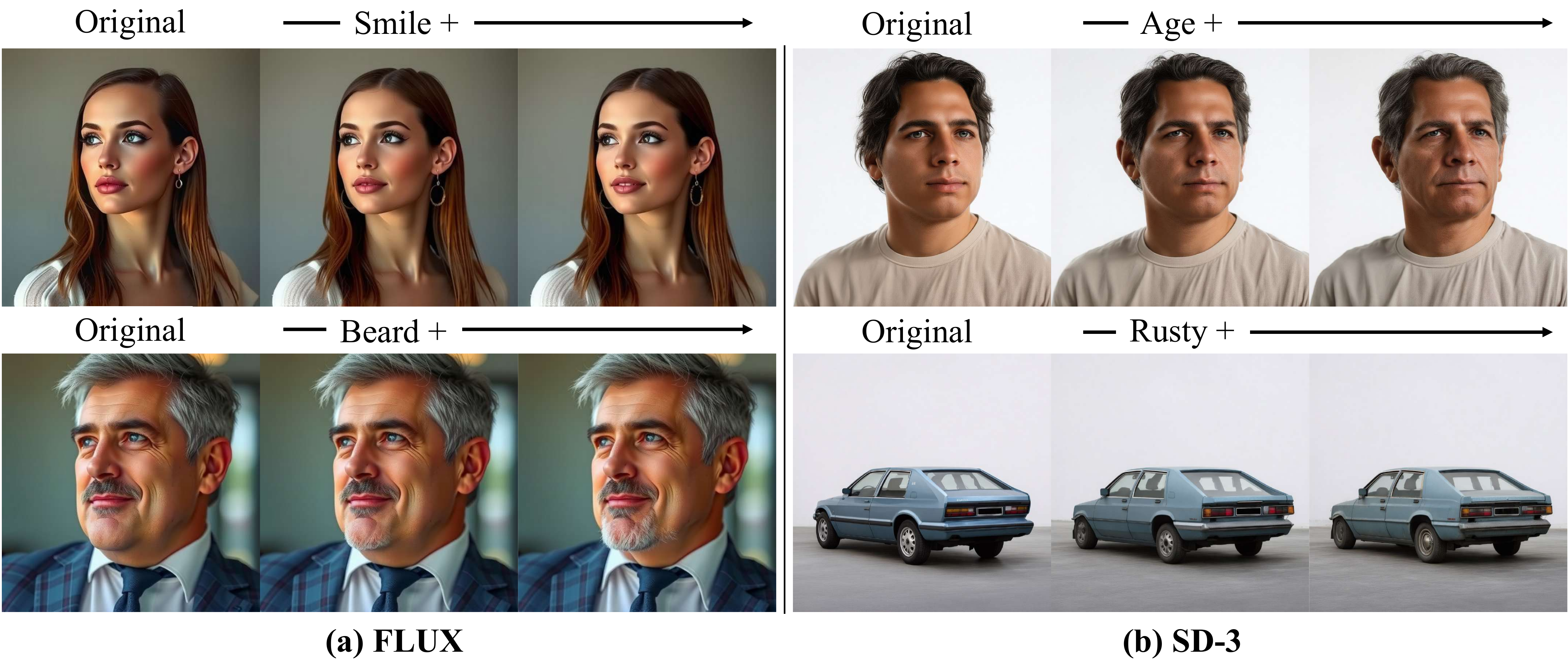}
    \caption{\textbf{Zero-shot generalization to FLUX and SD-3.} Text Slider can be directly applied to transformer-based diffusion models such as FLUX.1-schnell~\cite{flux2024} and SD-3~\cite{esser2024scaling} without retraining, further demonstrating the strong generalizability of our method.}
    \label{fig:sd-3-and-flux}
    \vspace{-1em}
\end{figure}

\begin{table}
  \resizebox{\columnwidth}{!}{
  \begin{tabular}{cccc}
    \toprule
    \textbf{Text-to-Image, SD-XL} & Effectiveness & Smoothness & Preservation \\
    \midrule
    Concept Slider~\cite{gandikota2024concept} & 3.44 & 3.27 & 3.01 \\
    Attribute Control~\cite{baumann2025attributecontrol} & 4.04 & 3.76 & 3.25 \\
    Text Slider (Ours) & \textbf{4.28} & \textbf{4.19}  & \textbf{3.85} \\
    \midrule
    \textbf{Text-to-Image, SD-1.5} & Effectiveness & Smoothness & Preservation \\
    \midrule
    Concept Slider~\cite{gandikota2024concept} & 4.14 & 3.81 & \textbf{3.98} \\
    Attribute Control~\cite{baumann2025attributecontrol} & 3.54 & 3.19 & 3.44 \\
    Text Slider (Ours) & \textbf{4.25} & \textbf{4.09} & 3.54 \\
    \midrule
    \textbf{Text-to-Video} & Effectiveness & Smoothness & Preservation \\
    \midrule
    Prompting~\cite{guo2023animatediff} & 3.51 & 2.47 & 2.14 \\
    Concept Slider~\cite{gandikota2024concept} & 3.84 & 3.80 & 3.56 \\
    Attribute Control~\cite{baumann2025attributecontrol} & 3.46 & 3.57 & \textbf{3.87} \\
    Text Slider (Ours) &  \textbf{4.02} & \textbf{3.94} & 3.84\\
    \midrule
    \textbf{Video-to-Video} & Effectiveness & Smoothness & Preservation \\
    \midrule
    Video-P2P~\cite{liu2024video} & 3.27 & 2.30 & 1.95 \\
    Concept Slider~\cite{gandikota2024concept} & 3.32 & 3.54 & \textbf{3.84} \\
    Attribute Control~\cite{baumann2025attributecontrol} & 3.28 & 3.55 & 3.73 \\
    Text Slider (Ours) & \textbf{3.69} & \textbf{3.82} & \textbf{3.84} \\
    \bottomrule
  \end{tabular}
  }

  \caption{\textbf{User Study.} Text Slider demonstrates superior performance across multiple diffusion models (\eg, SD-1.5 and SD-XL) and generation tasks (\eg, text-to-image, text-to-video, and video-to-video), evaluated across three criteria. We report the average absolute scores (scale of 1 to 5) across three questions per set. Higher scores indicate better performance.} 
  \label{table:user-study}
  \vspace{-1em}
\end{table}

\subsection{Generalization}
\label{sec:generalization}
\noindent\textbf{Transformer-based diffusion models.}
We further examine the generalizability of our method. Since Text Slider fine-tunes the appended LoRA modules on CLIP and OpenCLIP, it can be seamlessly applied to other text-guided diffusion models that also employ these text encoders. For instance, SD-3~\cite{esser2024scaling} incorporates CLIP, OpenCLIP, and T5~\cite{T5}; we apply Text Slider to its CLIP and OpenCLIP components. Likewise, FLUX.1-schnell~\cite{flux2024} utilizes CLIP and T5, where we inject the CLIP component of Text Slider into its CLIP encoder. In Figure~\ref{fig:sd-3-and-flux}, Text Slider adapts directly to these transformer-based diffusion models without retraining while maintaining the continuous control capability, highlighting its strong adaptability.

\begin{figure}[t]
    \centering
    \includegraphics[width=1.0\columnwidth]{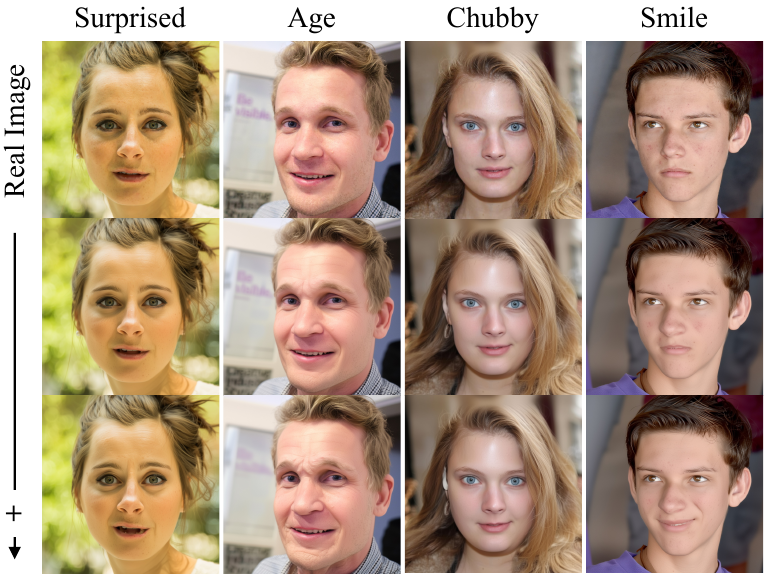}
    \caption{
\textbf{Results on Real Image Editing.} By inverting real images with ReNoise~\cite{garibi2024renoise} and applying our method, we achieve fine-grained attribute control on real images.}
    \label{fig:real-image-editing}
\end{figure}

\noindent\textbf{Real Image Editing.}
To evaluate real image editing, we integrate our method with ReNoise~\cite{garibi2024renoise} by inverting real images and regenerating them with specific attributes. As shown in Figure~\ref{fig:real-image-editing}, this demonstrates the effectiveness of our method for real image editing. We use SD-XL as the base model, set the guidance scale to 1.0, and disable the sliders until timestep $t = 550$ to better preserve consistency with the input image.

\subsection{Ablation Study}
\label{sec:ablation}
\noindent\textbf{Diffusion Noise Prediction.}
We evaluate the use of diffusion noise prediction as the training objective on SD-XL by fine-tuning the same LoRA modules in the text encoder and compare it with our default setting. Unlike our approach, noise prediction operates in the image space and requires backpropagation through the diffusion model, incurring significantly higher computational costs. As shown in Table~\ref{table:ablation}, our method achieves comparable $\Delta$CLIP and LPIPS performance while substantially reducing training time and GPU memory usage.

\begin{table}
  \resizebox{\columnwidth}{!}{
  \begin{tabular}{cccccc}
    \toprule
      & Time (s) & Mem. (GB) & \#Params(M) & $\Delta$CLIP ($\uparrow$) & LPIPS ($\downarrow$) \\
    \midrule
    Text Slider (Ours) & \textbf{550.59} & \textbf{5.68} & 1.53 & \textbf{1.285} & 0.022 \\
    Text Slider$^\dagger$ & 2295.91 & 12.94 & 1.53 & 0.859 & \textbf{0.017} \\
    \bottomrule
   \end{tabular}}
  \caption{\textbf{Comparison with Backpropagting though Diffusion Model.} Text Slider$^\dagger$ denotes the variant trained with diffusion noise prediction as the objective. Our results show that Text Slider achieves competitive performance without backpropagating through the diffusion model, significantly reducing computational costs, providing a 4$\times$ speedup in training and reducing GPU memory consumption by 2$\times$ compared to the diffusion-based variant.}
  \label{table:ablation}
\end{table}

\begin{table}
  \resizebox{\columnwidth}{!}{
    \begin{tabular}{lccccc}
        \toprule
        \textbf{CLIP} & \textbf{OpenCLIP} & Time (s) & Mem. (GB) & \#Params(M) \\
        \midrule
        \checkmark & \checkmark & 550.59 & 5.68 & 1.53\\
        \checkmark & \space & 215.50 & 1.32 & 0.28 \\
        \space & \checkmark & 505.79 & 4.76 & 1.25 \\
        \bottomrule
    \end{tabular}}
  \caption{\textbf{Comparison on Training Efficiency across Text Encoders.} Training on just one encoder (CLIP or OpenCLIP) reduces training time, GPU consumption, and parameters, while performance drops in $\Delta$CLIP and LPIPS stay minor and acceptable.}
  \label{table:text-encoder-ablation}
\end{table}

\begin{figure}[t]
    \centering
    \includegraphics[width=1.0\columnwidth]{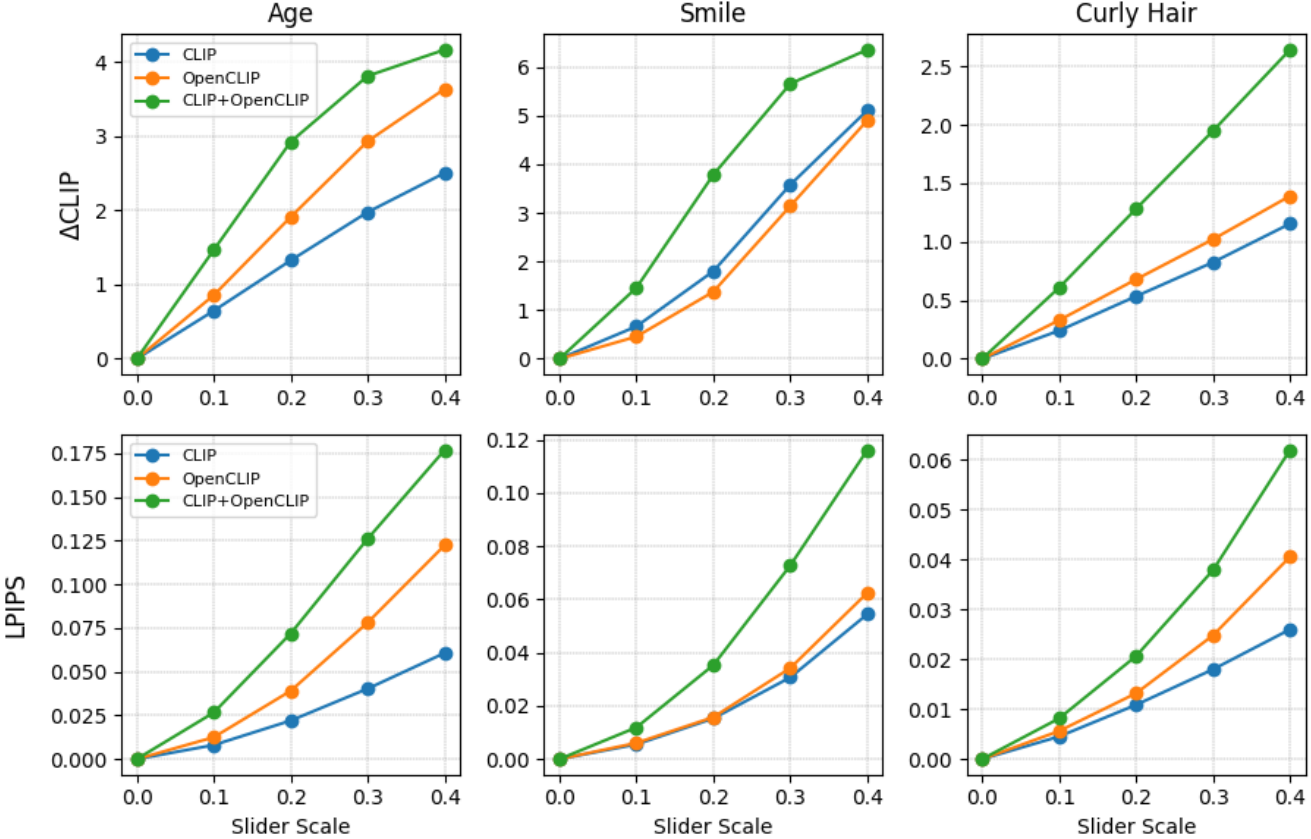}
    \caption{\textbf{Comparison on Performance across Text Encoders.} Training with a single text encoder (CLIP or OpenCLIP) results in reduced $\Delta$CLIP scores and LPIPS but remains within an acceptable range, providing a more training-efficient alternative.}
    \label{fig:clip-ablation}
\end{figure}

\noindent\textbf{CLIP Text Encoder.}
We also explore training our method solely on either CLIP or OpenCLIP under the rank-4 setting to explore more training-efficient strategies. Using a single text encoder and evaluating on SD-XL yields lower $\Delta$CLIP scores and LPIPS but remains within an acceptable range, offering a more efficient alternative, as shown in Table~\ref{table:text-encoder-ablation} and Figure~\ref{fig:clip-ablation}. This demonstrates Text Slider’s adaptability to smaller text encoders, enabling faster custom slider training with lower GPU requirements and scalability to larger concept sets, making it accessible to a broader range of users.

\begin{figure}[t]
    \centering
    \includegraphics[width=1.0\columnwidth]{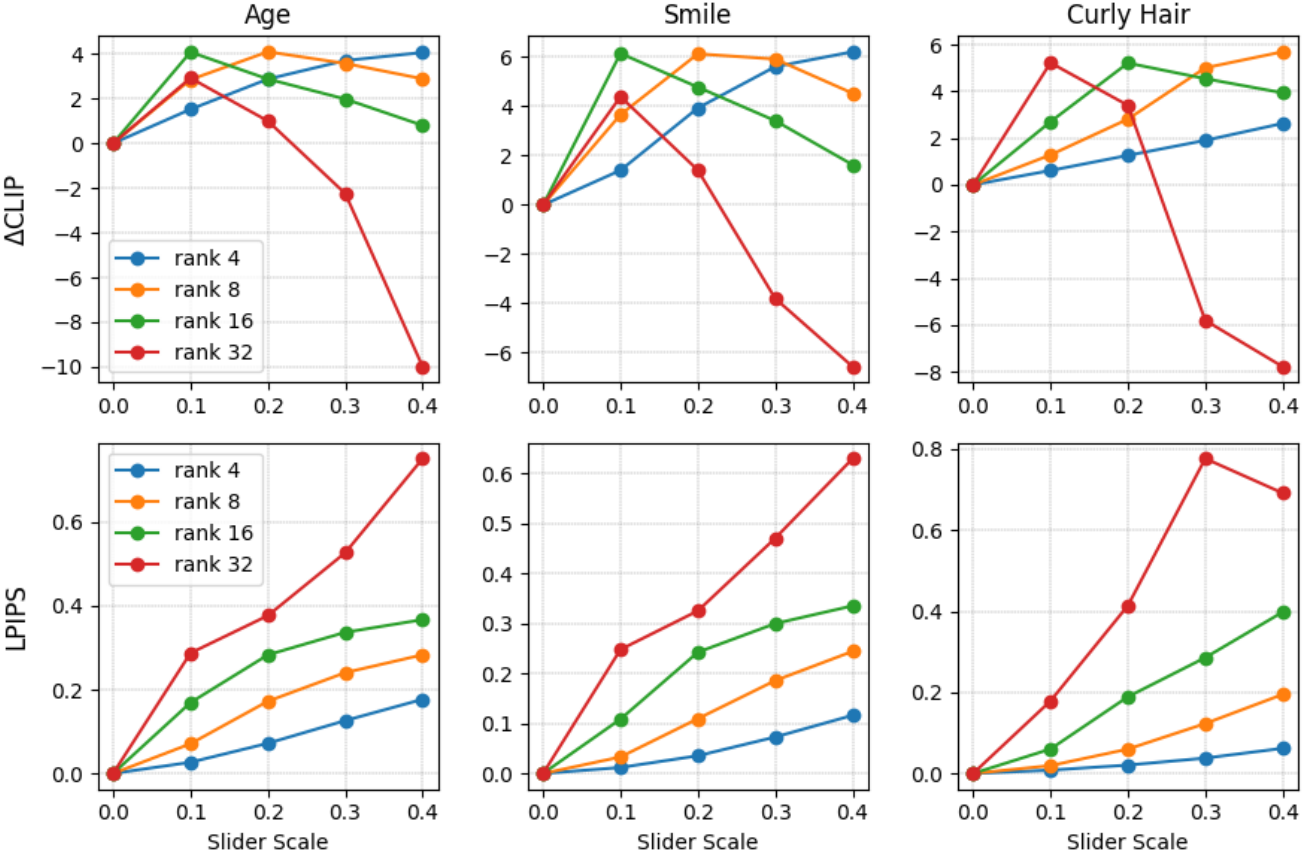}
    \caption{
\textbf{Rank Ablation.} Higher ranks often cause abrupt drops in $\Delta$CLIP scores beyond certain scales and increased LPIPS, narrowing the effective range of scaling factors. The low-rank (rank-4) setting offers a better balance of performance and stability.}
    \label{fig:rank-ablation}
    \vspace{-0.5em}
\end{figure}

\noindent\textbf{Rank Selection.}
We compared our default rank-4 setting with higher-rank (8, 16, 32). As shown in Figure~\ref{fig:rank-ablation}, increasing the rank often causes abrupt drops in $\Delta$CLIP scores beyond certain scales and higher LPIPS, narrowing the effective range of scaling factors and raising the risk of numerical underflow. For instance, the effective range for the age attribute shrinks from 0–0.4 at rank-4 to 0–0.2 at rank-8 and 0–0.1 at rank-16. In contrast, our low-rank setting achieves a more favorable balance between performance, efficiency, stability and usability.

\section{Conclusion}
We propose Text Slider, a continuous concept control method that is highly efficient, adaptable, plug-and-play, and composable. Our method significantly reduces the training time and GPU memory required to learn a slider. Moreover, it generalizes across different diffusion model architectures without retraining, whereas previous work either requires separate training for each model or time-consuming backpropagation through the diffusion model. Additionally, our approach naturally extends to text-to-video and video-to-video generation, enabling continuous concept control in both image and video domains.

\section{Acknowledgment}
This research is supported by National Science and Technology Council (NSTC) under the grant of NSTC-114-2634-F-002-004, NSTC-113-2634-F-002-008, NSTC-114-2221-E-001-016 and NSTC-114-2221-E-001-004 and Academia Sinica under the grant of AS-CDA-110-M09 and AS-IAIA-114-M10. 
{
    \small
    \bibliographystyle{ieeenat_fullname}
    \bibliography{main}
}

\newpage
\clearpage
\appendix
\setcounter{page}{1}
\setcounter{table}{0}
\renewcommand{\thetable}{A\arabic{table}}
\setcounter{figure}{0}    
\renewcommand{\thefigure}{A\arabic{figure}}
\setcounter{equation}{0}    
\renewcommand{\theequation}{A\arabic{equation}}
\maketitlesupplementary

\section{Limitation}
\label{sec:limitation}
Text Slider provides a training-efficient approach for continuous attribute modulation, enabling faster custom slider training with lower GPU requirements and scalability to larger concept sets, making it accessible to a broader range of users. However, our method inherits the limitation of low-rank adaptation~\cite{hu2022lora}, it remains sensitive to excessively large inference-time scaling factors, which can cause catastrophic forgetting of base knowledge and resulting in unnatural expressions or structural distortions.

\section{More Implementation Details}
\label{sec:supp-implementation}
\noindent\textbf{Model Checkpoints.}
For text-to-image experiments, we use the official SD-XL model checkpoint\footnote{\href{https://huggingface.co/stabilityai/stable-diffusion-xl-base-1.0}{https://huggingface.co/stabilityai/stable-diffusion-xl-base-1.0}} consistently throughout all evaluations. For SD-1.5, we adopt a high-quality community model\footnote{\label{fn:1}\href{https://huggingface.co/SG161222/Realistic\_Vision\_V6.0\_B1\_noVAE}{https://huggingface.co/SG161222/Realistic\_Vision\_V6.0\_B1\_noVAE}} to enhance generation fidelity.
For text-to-video experiments, we primarily employ SD-1.5 community checkpoints\footref{fn:1}\footnote{\label{fn:2}\href{https://civitai.com/models/43331/majicmix-realistic}{https://civitai.com/models/43331/majicmix-realistic}}, selected for their compatibility and visual quality.
For video-to-video experiments, we use the same community checkpoint\footref{fn:2} as MeDM~\cite{chu2024medm} to ensure consistency across our evaluations.

\noindent\textbf{Training Prompt Pairs Generation.}
We follow the same prompt generation strategy as Concept Slider~\cite{gandikota2024concept}. Specifically, we use OpenAI GPT-4o with a predefined system prompt\footnote{\href{https://github.com/rohitgandikota/sliders/blob/main/GPT\_prompt\_helper.ipynb}{https://github.com/rohitgandikota/sliders/GPT\_prompt\_helper.ipynb}}, which generates contrastive prompt pairs for the target concept.
More detailed examples of the training prompt pairs are provided in Table~\ref{table:detail-prompts}.

\section{Qualitative Comparison}
\label{sec:supp-qualitative}
\begin{figure*}[t]
    \centering
    \includegraphics[width=0.9\linewidth]{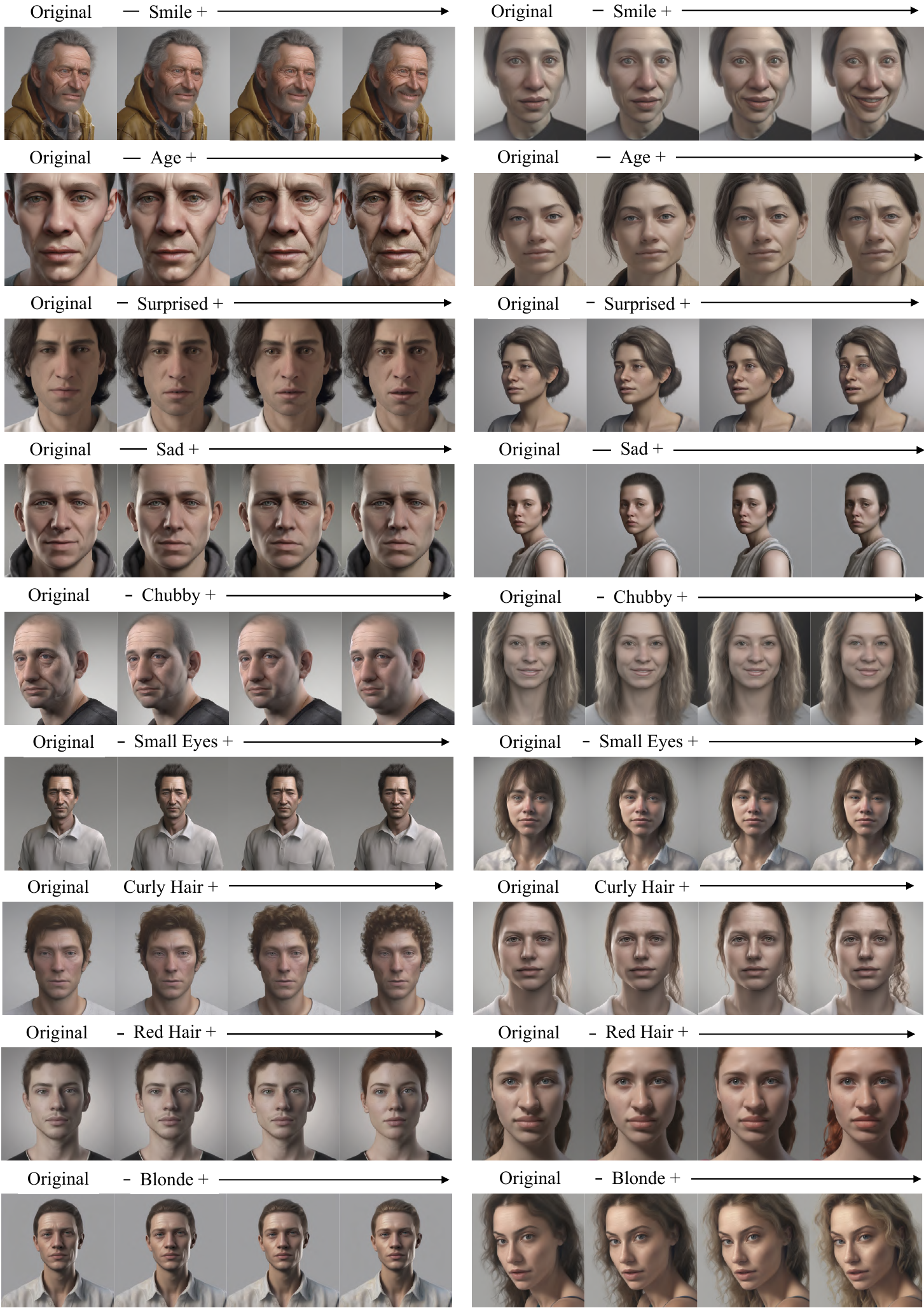}
    \caption{
\textbf{More Qualitative Results on SD-XL.} We present more results using Text Slider across face, eyes and hair-related attributes.}
    \label{fig:more-results}
\end{figure*}

\begin{figure*}[t]
    \centering
    \includegraphics[width=0.9\linewidth]{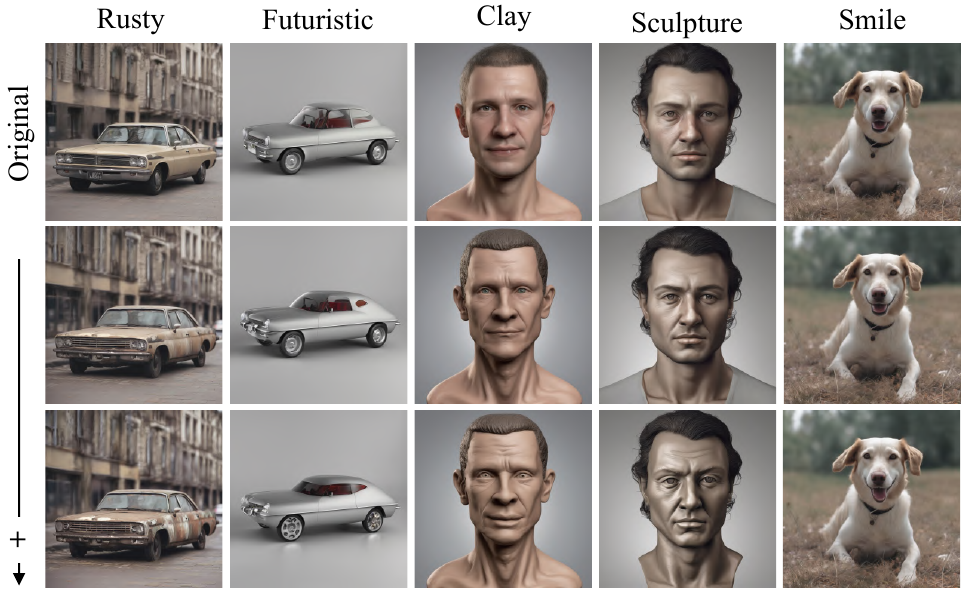}
    \caption{
\textbf{More Qualitative Results on SD-XL.} Our method is also effective for attributes related to cars, styles, and dogs.}
    \label{fig:more-results-2}
\end{figure*}

\begin{figure*}[t]
    \centering
    \includegraphics[width=0.9\linewidth]{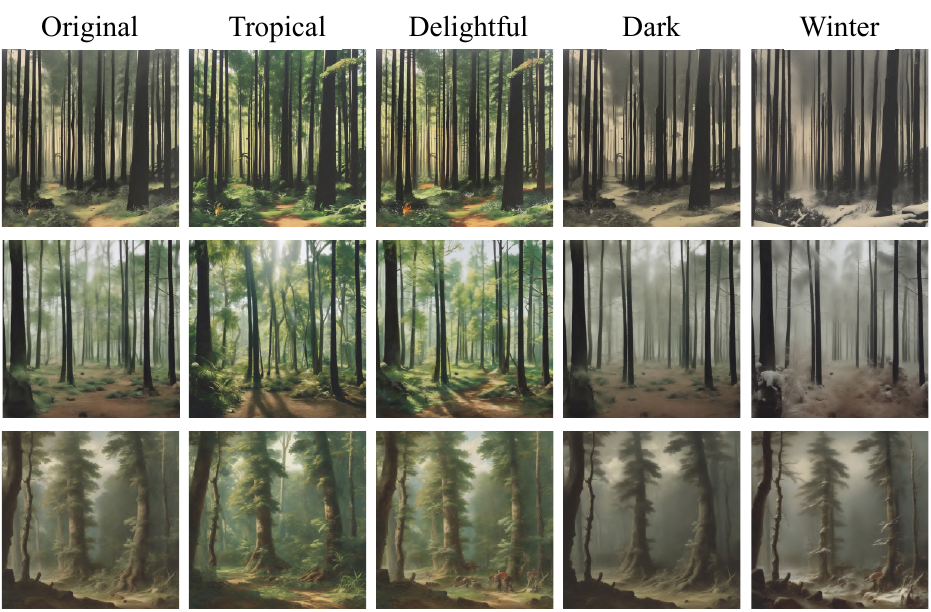}
    \caption{
\textbf{Qualitative Results for scene-related attributes on SD-XL.}}
    \label{fig:more-results-scene}
\end{figure*}

\begin{figure*}[t]
    \centering
    \includegraphics[width=0.82\linewidth]{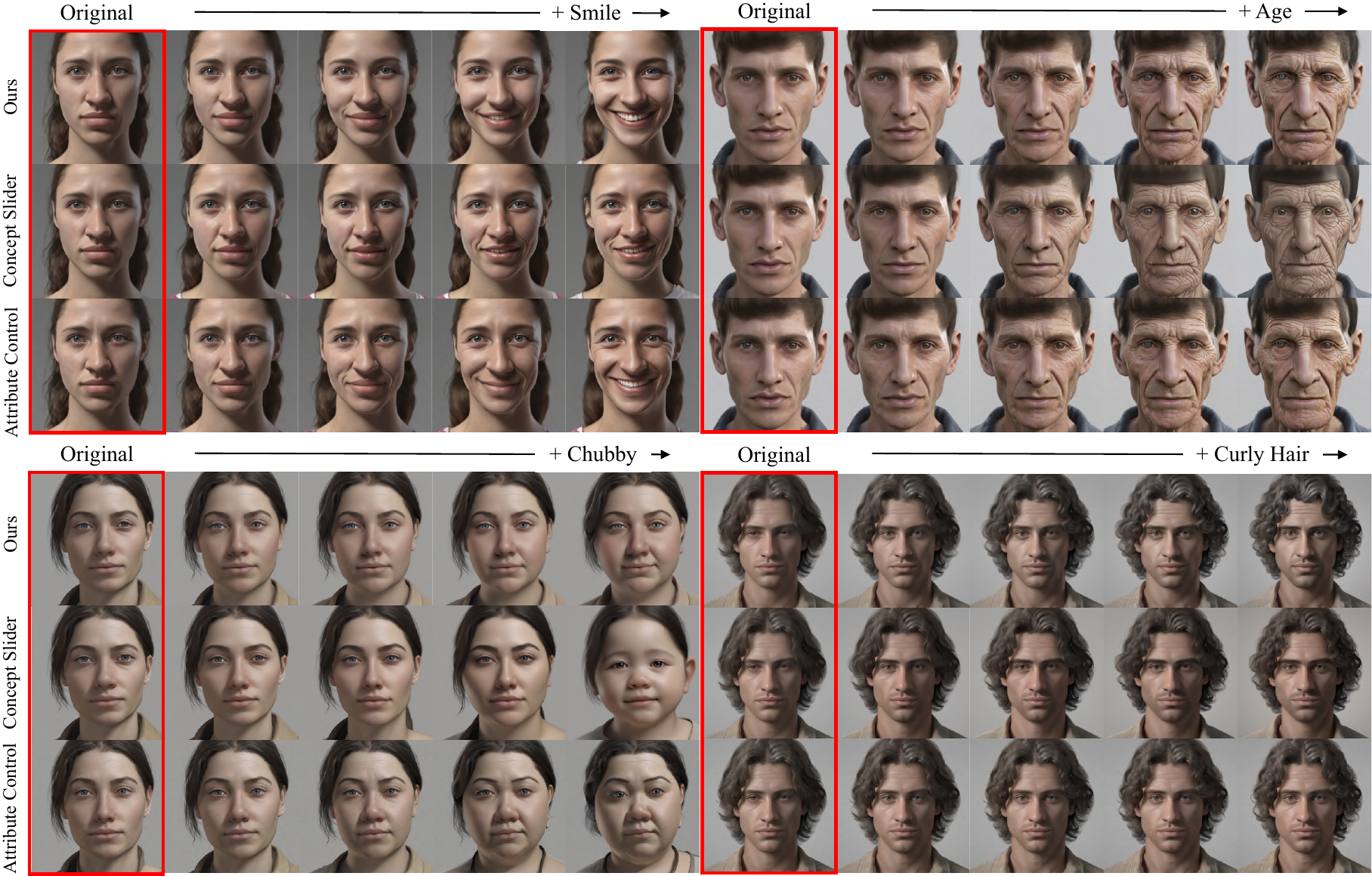}
    \caption{
\textbf{Qualitative Comparison of Text-to-Image Results on SD-XL.} We qualitatively compare Text Slider with Concept Slider~\cite{gandikota2024concept} and Attribute Control~\cite{baumann2025attributecontrol} on SD-XL~\cite{podell2023sdxl} across four attributes, smile, age, chubby and curly hair. Each attribute evaluated at four levels of intensity. Red boxes highlight the original generated images for reference.}
    \label{fig:t2i-comparison}
\end{figure*}

\begin{figure*}[t]
    \centering
    \includegraphics[width=0.82\linewidth]{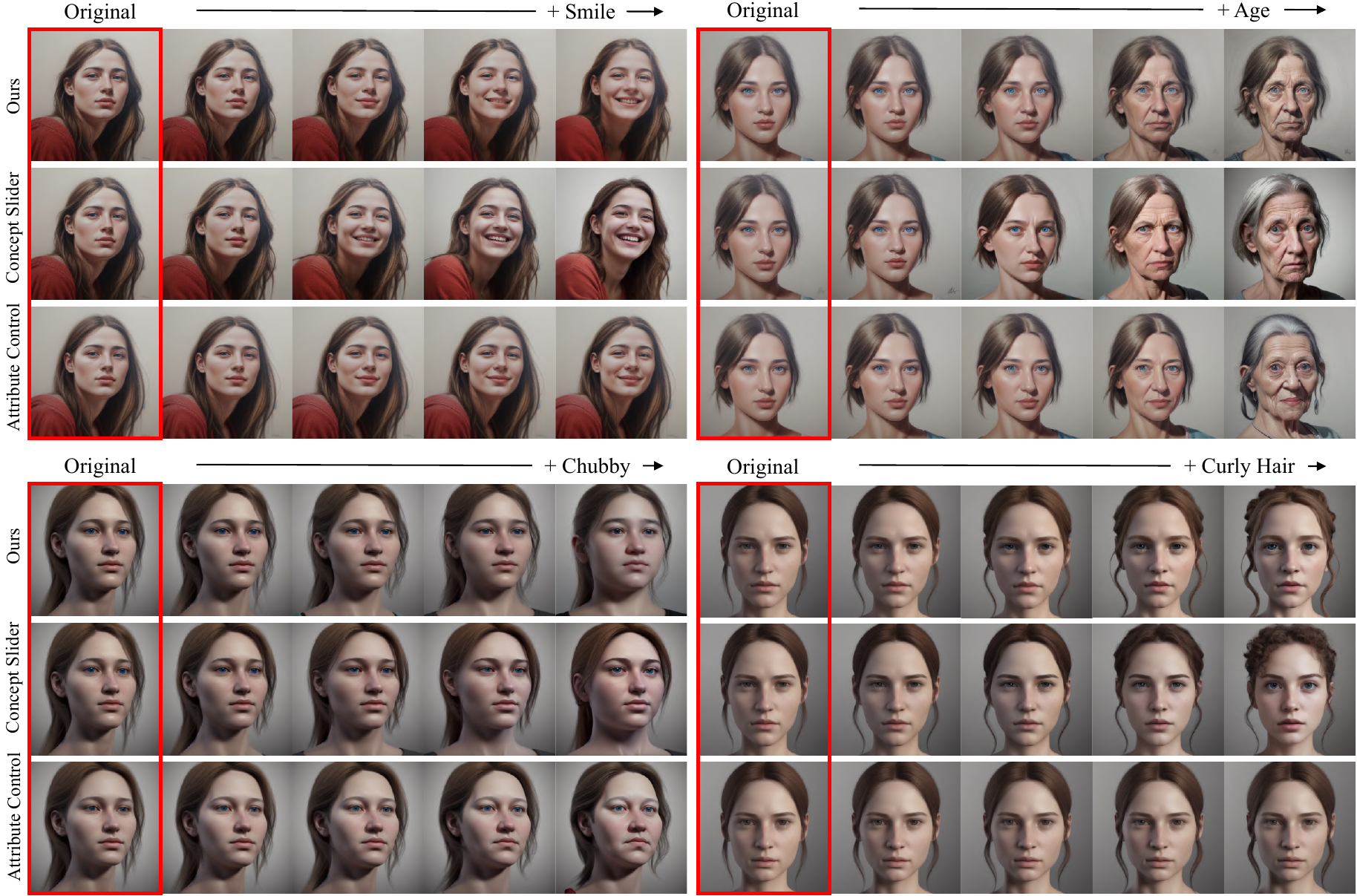}
    \caption{
\textbf{Qualitative Comparison of Text-to-Image Results on SD-1.} We qualitatively compare Text Slider with Concept Slider~\cite{gandikota2024concept} and Attribute Control~\cite{baumann2025attributecontrol} on SD-1~\cite{rombach2022high} across four attributes, smile, age, chubby and curly hair. Each attribute evaluated at four levels of intensity. Red boxes highlight the original generated images for reference.}
    \label{fig:t2i-comparison-sd1}
\end{figure*}

\noindent\textbf{Text-to-Image Generation.}
Figure~\ref{fig:more-results}-\ref{fig:more-results-scene} present more diverse results on SD-XL using Text Slider. Figure~\ref{fig:t2i-comparison} and \ref{fig:t2i-comparison-sd1} provides additional qualitative comparisons with baseline methods on SD-XL and SD-1, respectively. Notably, unlike Concept Slider, which requires model-specific training, Text Slider generalizes across different architectures that share the same text encoder without the need for retraining and achieves comparable results. 

\begin{figure*}[t]
    \centering
    \includegraphics[width=0.85\linewidth]{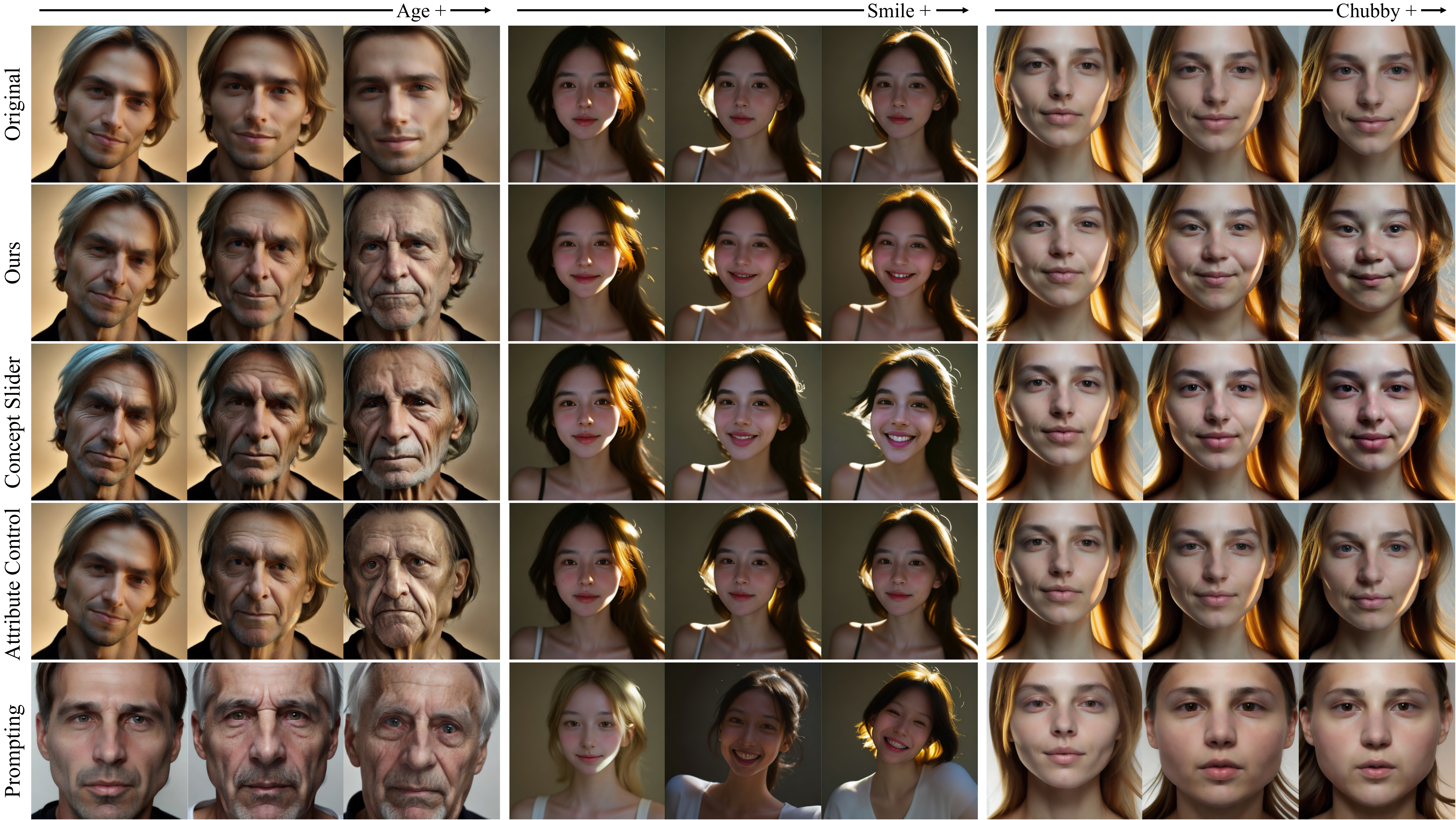}
    \caption{
\textbf{Qualitative Comparison of Text-to-Video Results.} We compare AnimateDiff~\cite{guo2023animatediff} integrated with Text Slider, Concept Slider~\cite{gandikota2024concept}, and Attribute Control~\cite{baumann2025attributecontrol} across three attributes. For each video, three representative frames are sampled to illustrate the gradual progression of attribute intensity over time.}
    \label{fig:t2v-comparison}
    \vspace{-0.3em}
\end{figure*}

\noindent\textbf{Text-to-Video Generation.}
Figure~\ref{fig:t2v-comparison} showcase a detailed comparison with baseline methods. To ensure a fair comparison, we primarily focus on person-related attributes, as baseline methods like Attribute Control have limited ability to manipulate global properties such as style or weather.

\begin{figure*}[t]
    \centering
    \includegraphics[width=0.85\linewidth]{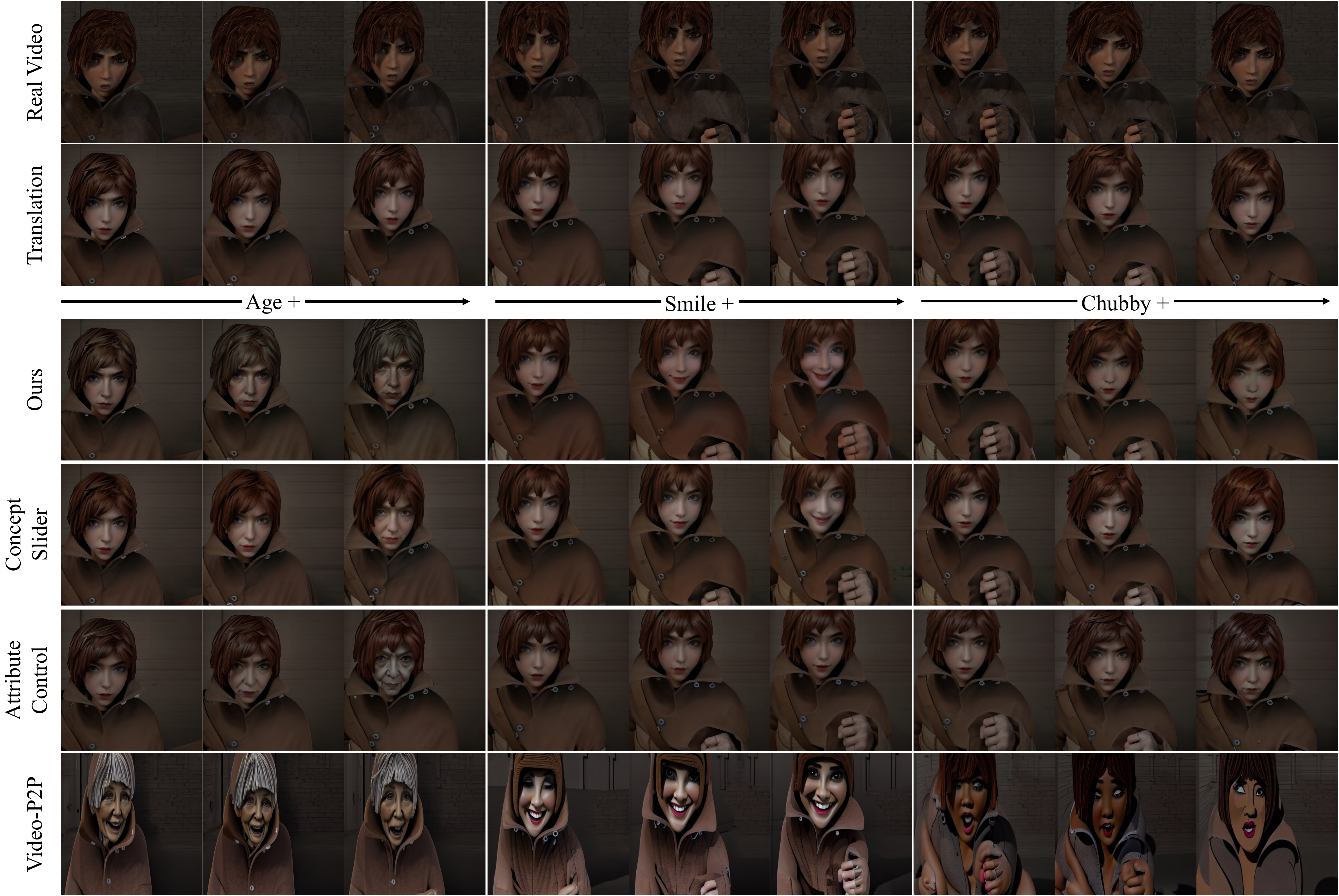}
    \caption{
\textbf{Qualitative Comparison of Video-to-Video Results.} Real videos are first translated using MeDM~\cite{chu2024medm} with SDEdit~\cite{meng2022sdedit}, followed by editing with Text Slider, Concept Slider~\cite{gandikota2024concept}, and Attribute Control~\cite{baumann2025attributecontrol} across three attributes. For Video-P2P, real videos are first inverted and then edited using attention map-based control. For each video, three representative frames are sampled to illustrate the gradual progression of attribute intensity over time.}
    \label{fig:v2v-comparison}
\end{figure*}

\noindent\textbf{Video-to-Video Generation.}
As shown in the baseline comparison in Figure~\ref{fig:v2v-comparison}, our approach delivers competitive visual quality with significantly lower computational overhead. In contrast, Video-P2P struggles with subtle facial edits and often introduces artifacts, while also requiring per-video model tuning. Text Slider, by comparison, offers a plug-and-play solution that generalizes across videos without additional fine-tuning.

\section{User Study}
\label{sec:supp-user-study}
\begin{figure*}[t]
    \centering
    \includegraphics[width=1.0\linewidth]{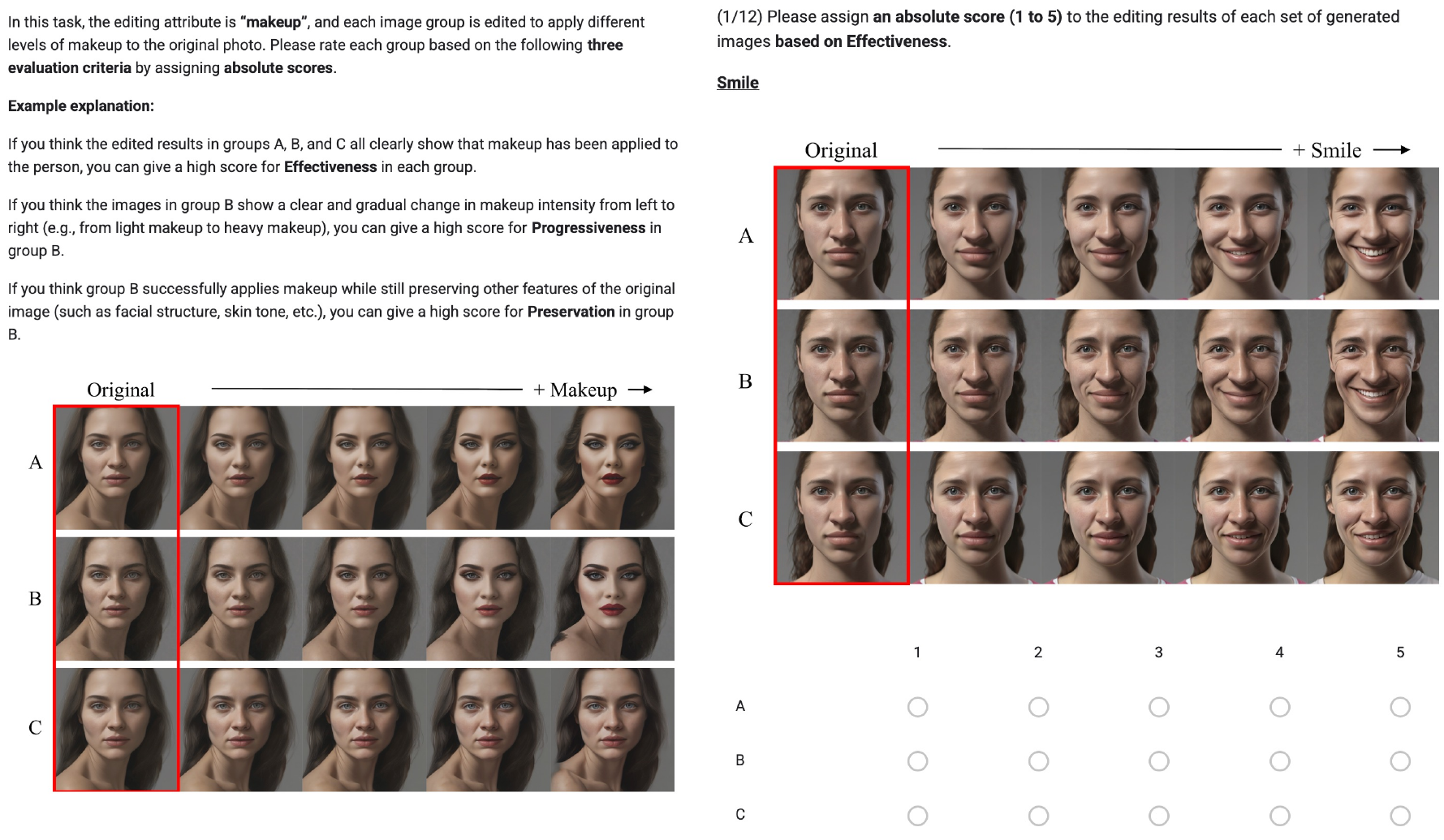}
    \caption{
\textbf{User Study Example.} Instructions and an example question on the left are provided to make evaluators familiar with the criteria, while the actual evaluation on the right presents three rows each denotes different methods, with their order randomized to prevent bias and ensure a fair assessment.}
    \label{fig:user-study-example}
\end{figure*}

In Figure~\ref{fig:user-study-example}, we illustrate some sample questions of the questionnaire in our user study. An instruction and an example question are provided on the left to let evaluators familiar with the criteria and question format before starting the actual evaluation on the right. Each question consists of three rows corresponding to different methods, with their order randomized to prevent bias and ensure fairness in assessment.

\section{Ablation Study}
\label{sec:supp-ablation}
\begin{figure*}[t]
    \centering
    \includegraphics[width=1.0\linewidth]{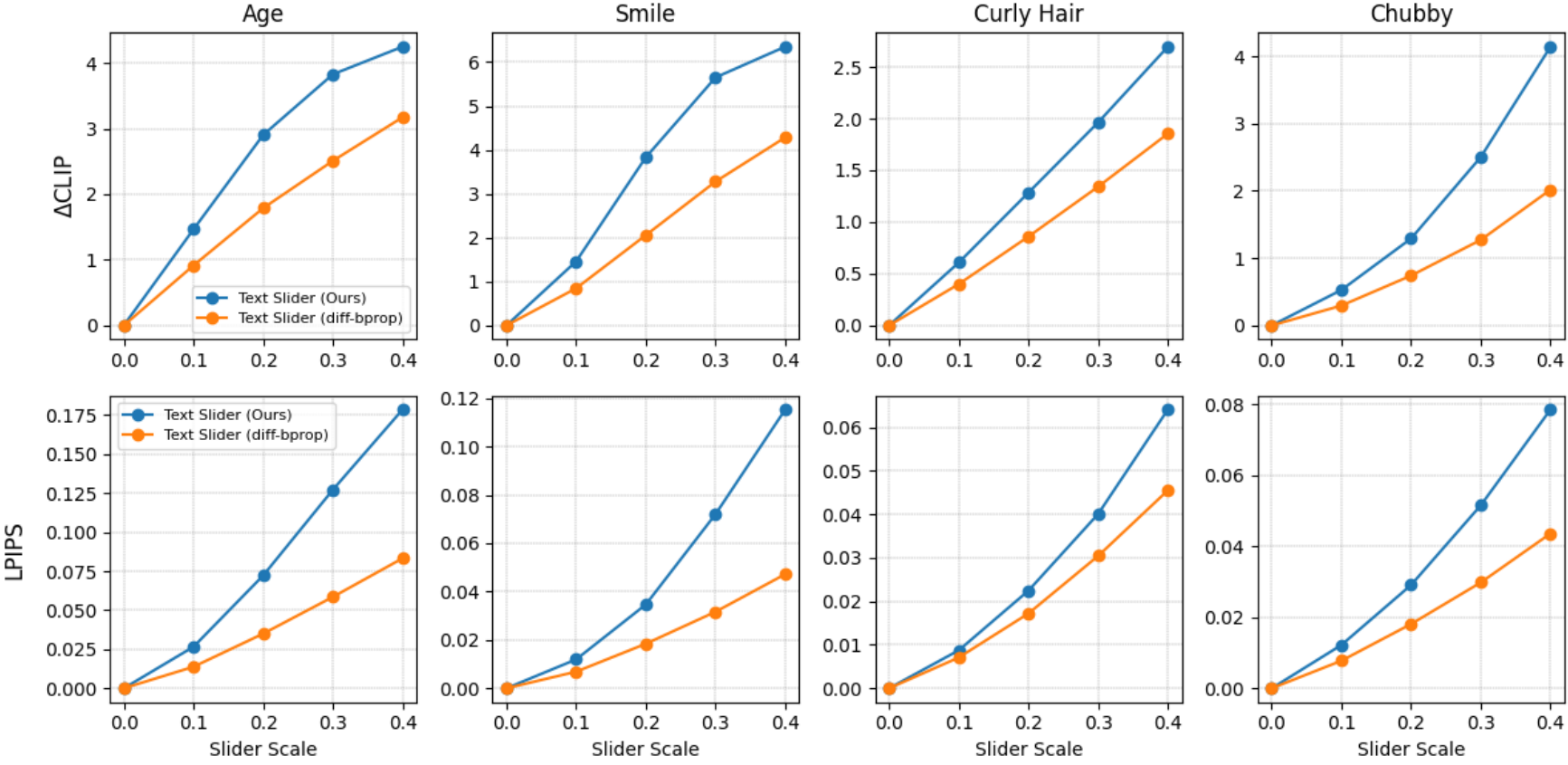}
    \caption{
\textbf{Comparison on Performance with Diffusion Noise Prediction.} \texttt{diff-bprop} denotes the setting where the same LoRA modules are injected into the text encoder, but backpropagation is performed through the diffusion model. Our method achieves comparable performance in both $\Delta$CLIP and LPIPS.}
    \label{fig:diff-bprop-curve}
\end{figure*}

\begin{figure*}[t]
    \centering
    \includegraphics[width=1.0\linewidth]{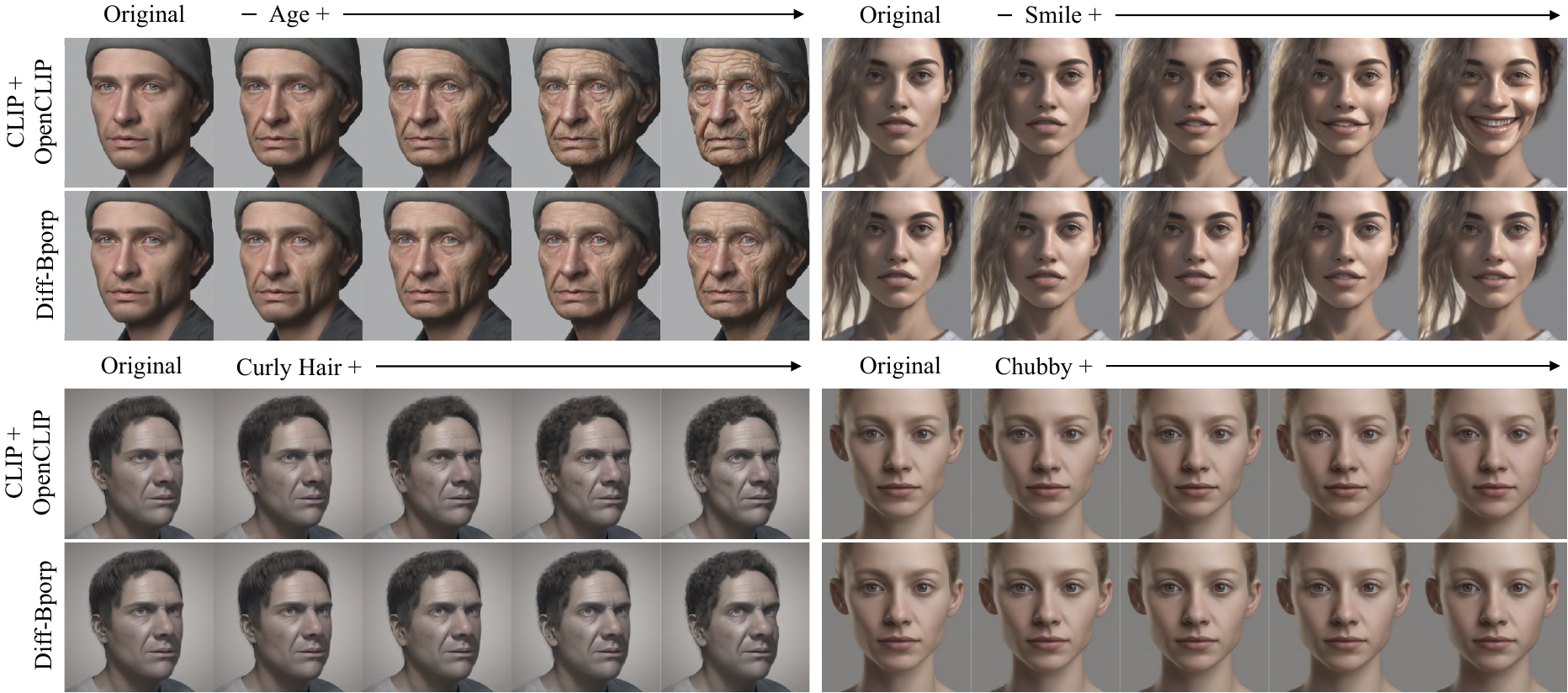}
    \caption{
\textbf{Qualitative Comparison on Diffusion Noise Prediction.} We evaluate four attributes—age, smile, curly hair, and chubby—using five levels on a 0–0.4 scale with 0.1 intervals. Our method achieves comparable qualitative performance to the variant that backpropagates through the diffusion model, while significantly reducing computational costs.}
    \label{fig:diff-bprop-qual-comparison}
\end{figure*}

\noindent\textbf{Diffusion Noise Prediction.}
In Figure~\ref{fig:diff-bprop-curve}, we compare the diffusion noise prediction setting by reporting $\Delta$CLIP and LPIPS across five attribute intensities for four attributes: age, smile, curly hair, and chubby. Figure~\ref{fig:diff-bprop-qual-comparison} further provides qualitative results, confirming that our method achieves performance comparable to the setting that backpropagates through the diffusion model.

\begin{figure*}[t]
    \centering
    \includegraphics[width=1.0\linewidth]{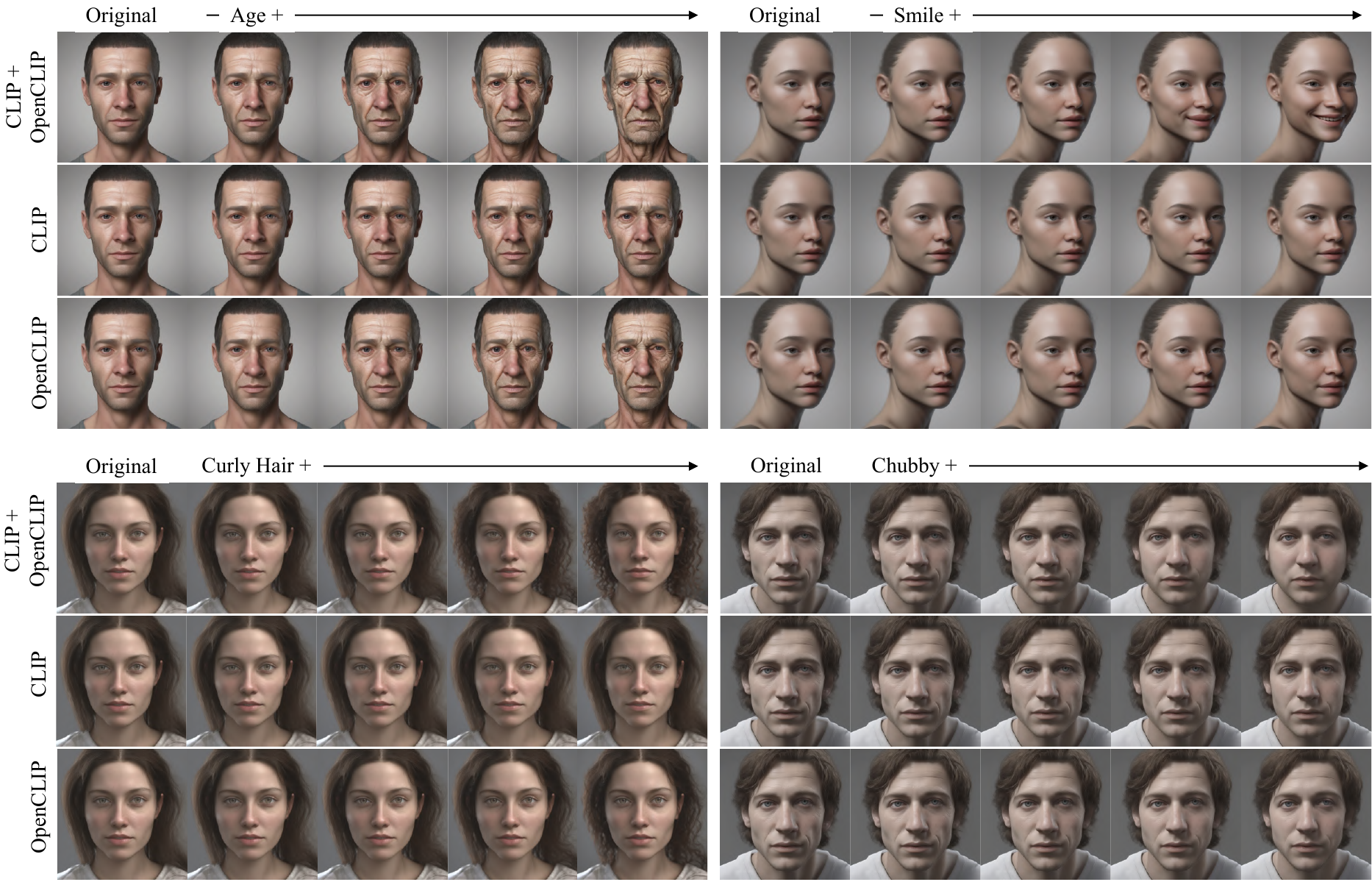}
    \caption{
\textbf{Qualitative Comparison across CLIP Text Encoders.} All settings enable effective attribute manipulation, with single text encoder configurations offering a more training-efficient alternative. However, the default configuration provides stronger control over certain attributes (\eg, curly hair, chubby), enabling a broader and more diverse range of concepts.}
    \label{fig:clip-ablation-qual-comparison}
\end{figure*}

\noindent\textbf{CLIP Text Encoder.}
We present a qualitative comparison of three settings in Figure~\ref{fig:clip-ablation-qual-comparison}: our default (CLIP+OpenCLIP), CLIP-only, and OpenCLIP-only. All settings effectively manipulate attributes, but the default setting offers stronger control over certain attributes (\eg, curly hair, chubby), enabling a broader and more diverse range of concepts.

\section{Training-Free Text Slider}
\label{sec:training-free}
\begin{figure*}[t]
    \centering
    \includegraphics[width=1.0\linewidth]{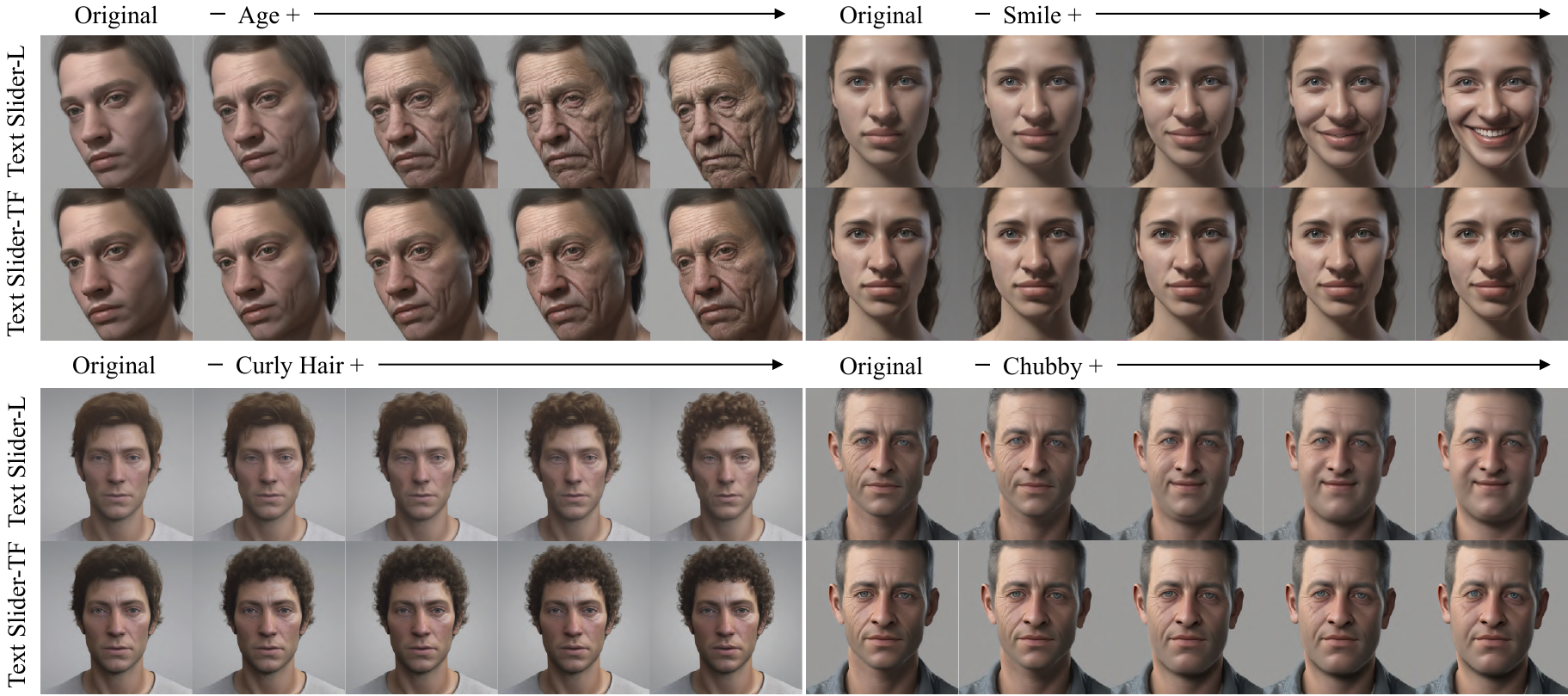}
    \caption{
\textbf{Qualitative Comparison of the LoRA-based and the Training-Free Results on SD-XL.} \texttt{Text Slider-L} denotes our LoRA-based method and \texttt{Text Slider-TF} indicates the training-free variant. We evaluate four attributes—age, smile, curly hair, and chubby—using five levels on a 0–0.4 scale with 0.1 intervals. Our LoRA-based method achieves competitive performance compared to the training-free variant, while exhibiting more pronounced visual editing effects and reducing inference time from three forward passes to a single forward pass.}
    \label{fig:lora-tf-figure}
\end{figure*}

\begin{figure*}[t]
    \centering
    \includegraphics[width=1.0\linewidth]{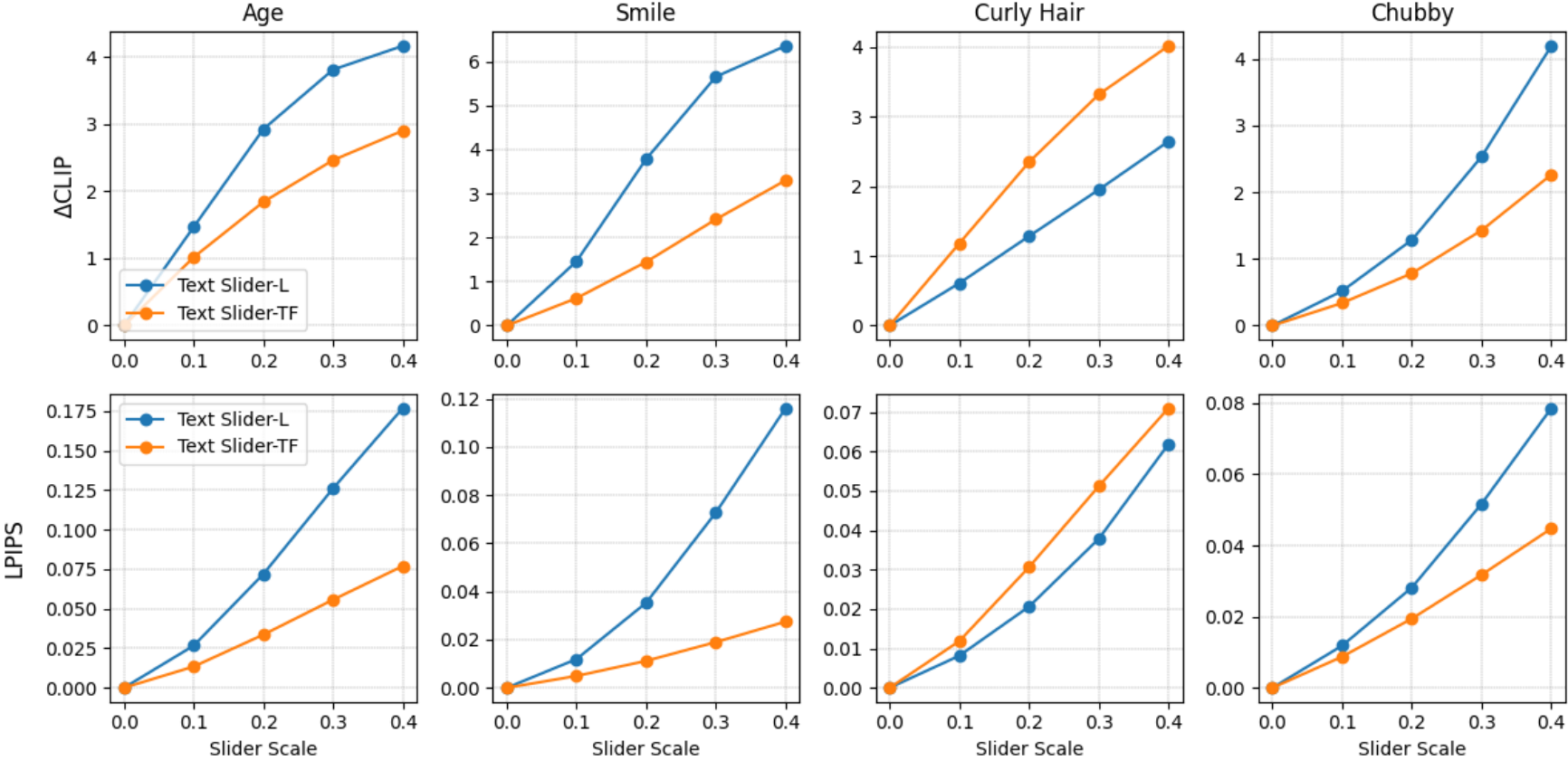}
    \caption{
\textbf{Quantitative Comparison of the LoRA-based and the Training-Free Results on SD-XL.} \texttt{Text Slider-L} denotes our LoRA-based method and \texttt{Text Slider-TF} indicates the training-free variant. We report performance metrics using $\Delta$CLIP and LPIPS across four attributes—\textit{age}, \textit{smile}, \textit{curly hair}, and \textit{chubby}—evaluated at five levels of attribute intensity (slider scales) from 0 to 0.4. Our LoRA-based method achieves competitive performance compared to the training-free variant, while exhibiting more pronounced visual editing effects and reducing inference time from three forward passes to a single forward pass.}
    \label{fig:lora-tf-curve}
\end{figure*}

To further investigate the fundamental mechanism of Text Slider, we explore a training-free version that leverages the near-linear properties of the text embedding space. Specifically, we test whether the directional shift identified by our framework can be applied directly via vector arithmetic without any parameter updates. This variant constructs modified text embeddings $\tau_{\mathrm{mod}}$ as follows:
\begin{equation}
\tau_{\mathrm{mod}} = \tau_{\theta}(c_t) + s \cdot [\tau_{\theta}(c_+) - \tau_{\theta}(c_-)],
\end{equation}
where $c_t, c_+$, and $c_-$ represent the target, positive, and negative prompts, respectively. This formulation serves as an empirical validation of our core hypothesis that continuous attribute control can be effectively achieved by navigating along specific directional axes in the text embedding space. \\
\indent For the comparison with LoRA-based Text Slider, we provide qualitative and quantitative evaluations in Figure~\ref{fig:lora-tf-figure} and \ref{fig:lora-tf-curve}. While the training-free variant achieves competitive performance, it reveals two primary limitations compared to our proposed framework. 
First, the training-free variant exhibits lower control responsiveness; under the same scaling range of $s$, the visual editing intensity is noticeably less pronounced than that achieved by the LoRA-based method. This suggests that the LoRA fine-tuning process successfully distills and amplifies the target concept, allowing for more significant attribute manipulation. Second, the training-free approach incurs a significant computational overhead during inference. It requires three separate forward passes through the text encoder to compute the modified embedding on-the-fly for each concept.
In contrast, our default LoRA-based method distills this directional knowledge into the model weights, requiring only a single forward pass at inference time. These results suggest that while the embedding space exhibits well-behaved near-linear characteristics, our LoRA framework provides a more responsive and inference-efficient solution for practical deployment.

\section{Societal Impact}
Text Slider offers an efficient approach for continuous attribute control in image and video synthesis, enabling creative applications in design and entertainment. Its efficiency makes advanced generative tools more accessible to users with limited computational resources. However, the ability to manipulate visual attributes raises risks of misuse in misinformation, deepfakes, and identity spoofing. Therefore, responsible deployment should include safeguards such as content provenance tracking, user consent mechanisms, and bias audits to ensure ethical and fair use.

\twocolumn[{
\renewcommand\twocolumn[1][]{#1}%
\begin{center}
\vspace{-2em}
\resizebox{\textwidth}{!}{
\begin{tabular}{|l|l|l|l|l|}
\hline
\textbf{Slider} & \textbf{Target} & \textbf{Positive} & \textbf{Negative} & \textbf{Preserved}\\
\hline
Age & \makecell{person} & \makecell{person, elderly, wrinkles, gray hair, \\aged skin, signs of maturity, wise expression} & \makecell{person, young, smooth skin, \\youthful appearance, no wrinkles, \\energetic expression} & \makecell{white race, black race, indian race, \\asian race, hispanic race ; male, female}\\
\hline
Smile & \makecell{person} & \makecell{person, smiling, happy face, big smile, \\joyful expression} & \makecell{person, frowning, grumpy, \\sad, neutral expression} & \makecell{white race, black race, indian race, \\asian race, hispanic race ; male, female}\\
\hline
Chubby & \makecell{person} &\makecell{person, chubby, round face, \\soft features, fuller cheeks, plump body shape} &  \makecell{person, skinny, thin face, \\sharp features, slim body shape} & \makecell{white race, black race, indian race, \\asian race, hispanic race ; male, female}\\
\hline
Beard & \makecell{person} & \makecell{person, full beard, thick facial hair, \\well-groomed beard, masculine appearance} & \makecell{person, clean-shaven, no facial hair, \\smooth chin, youthful look} & \makecell{white race, black race, indian race, \\asian race, hispanic race ; male, female}\\
\hline
Makeup & \makecell{person} & \makecell{person, wearing makeup, \\well-applied foundation, eye shadow, \\lipstick, glamorous look, \\enhanced facial features} & \makecell{person, no makeup, natural skin, \\bare face, unaltered appearance} & \makecell{white race, black race, indian race, \\asian race, hispanic race ; male, female}\\
\hline
Small eyes & \makecell{person} & \makecell{person, small eyes, narrow eye shape, \\subtle eyelids, low eye-to-face ratio} & \makecell{person, large eyes, wide eye shape, \\prominent eyelids, high eye-to-face ratio} & \makecell{white race, black race, indian race, \\asian race, hispanic race ; male, female}\\
\hline
Surprised & \makecell{person} & \makecell{person, surprised expression, wide eyes, \\raised eyebrows, open mouth, shocked face, \\startled posture, expressive emotion} & \makecell{person, neutral expression, relaxed face, \\calm demeanor, closed mouth, steady gaze, \\no visible emotion} & \makecell{white race, black race, indian race, \\asian race, hispanic race ; male, female}\\
\hline
Curly hair & \makecell{person} & \makecell{person, curly hair, defined curls, \\voluminous texture, spiral or wavy patterns, \\natural bounce, well-defined ringlets} & \makecell{person, straight hair, smooth texture, \\no curls or waves, flat appearance} & \makecell{white race, black race, indian race, \\asian race, hispanic race ; male, female ; \\short hair, long hair, medium length hair}\\
\hline
Red hair & \makecell{person} & \makecell{person, red hair, vibrant copper tones, \\fiery orange red shades, \\natural auburn highlights, \\bold striking hair color} & \makecell{person, blond hair, black hair, \\brown hair, grey hair, non-red shades} & \makecell{white race, black race, indian race, \\asian race, hispanic race ; male, female ; \\short hair, medium hair, long hair ; \\straight hair, wavy hair, curly hair}\\
\hline
Blonde & \makecell{person} & \makecell{person, blonde hair, \\golden tones, light yellow shades, \\bright and radiant hair color} & \makecell{person, dark hair, black hair, brown hair, \\red hair, non-blonde shades} & \makecell{white race, black race, indian race, \\asian race, hispanic race ; male, female ; \\short hair, medium hair, long hair ; \\straight hair, wavy hair, curly hair}\\
\hline
Cartoon & \makecell{} & \makecell{cartoon style, exaggerated features, \\bold outlines, flat shading, \\vibrant colors, stylized characters, \\playful proportions, simplified textures, \\hand-drawn appearance} & \makecell{realistic style, natural proportions, \\detailed textures, realistic lighting, \\lifelike shading, photographic accuracy} & \makecell{white race, black race, indian race, \\asian race, hispanic race ; male, female ; \\urban background, nature background, \\indoor scene}\\
\hline
Pixar & \makecell{} & \makecell{pixar style, 3D animation, \\smooth and rounded features, \\expressive faces, high-quality rendering, \\vibrant and clean visuals, \\family-friendly aesthetic} & \makecell{realistic style, detailed textures, \\natural proportions, lifelike appearance, \\photographic realism} & \makecell{white race, black race, indian race, \\asian race, hispanic race ; male, female ; \\child, adult, elderly}\\
\hline
Clay & \makecell{} & \makecell{clay style, claymation look, sculpted textures, \\hand-molded appearance, matte surfaces, \\visible fingerprints, soft rounded edges, \\stop-motion aesthetic, playful handcrafted feel} & \makecell{realistic style, smooth digital textures, \\lifelike proportions, clean lines, \\high detail realism, photographic surfaces} & \makecell{white race, black race, indian race, \\asian race, hispanic race ; male, female ; \\indoor setting, outdoor setting, \\neutral background, colorful background}\\
\hline
Sculpture & \makecell{} & \makecell{sculpture, carved appearance, \\stone or marble texture, rigid posture, \\chiseled features, statue-like, solid material, \\classical sculpture aesthetic, matte surface} & \makecell{lifelike, soft skin, natural textures, \\fluid posture, organic materials, \\realistic surface detail} & \makecell{white race, black race, indian race, \\asian race, hispanic race ; male, female}\\
\hline
Tropical & \makecell{} & \makecell{tropical scene, lush green palm trees, \\warm sandy beaches, turquoise ocean, \\humid atmosphere, exotic plants, \\bright sunlight, vibrant flowers, \\tropical wildlife} & \makecell{non-tropical scene, dry landscape, \\temperate forest, rocky terrain, cool climate, \\muted colors, overcast sky} & \makecell{white race, black race, indian race, \\asian race, hispanic race ; male, female ; \\urban setting, nature background, \\indoor setting}\\
\hline
Winter & \makecell{} & \makecell{winter scene, snow-covered landscape, \\icy ground, frosty trees, frozen lakes, \\snowflakes falling, cloudy sky, \\cold atmosphere, winter lights, \\visible snow piles} & \makecell{summer scene, dry ground, \\green grass, leafy trees, clear sky, \\warm lighting, no snow, sunlit atmosphere} & \makecell{white race, black race, indian race, \\asian race, hispanic race ; male, female ; \\mountain setting, urban setting, countryside}\\
\hline
Delightful & \makecell{} & \makecell{delightful atmosphere, joyful expressions, \\cheerful colors, heartwarming scene, \\positive energy, vibrant and lively mood} & \makecell{gloomy atmosphere, sad expressions, \\dull colors, depressing scene, \\negative emotion, dark and heavy mood} & \makecell{white race, black race, indian race, \\asian race, hispanic race ; male, female ; \\indoor setting, outdoor setting, \\urban background, natural background}\\
\hline
Rusty & \makecell{car} & \makecell{car, rusty, corroded metal, peeling paint, \\oxidized surface, aged appearance, \\weathered condition} & \makecell{car, clean, polished, shiny surface, \\new paint, well-maintained, \\pristine condition} & \makecell{sedan, SUV, truck, convertible ; \\red, blue, black, white, silver}\\
\hline
Futuristic & \makecell{car} & \makecell{car, futuristic, sleek aerodynamic design, \\glowing neon lights, \\advanced technology features, \\metallic surfaces, sci fi style, \\high-tech appearance} & \makecell{car, traditional, old-fashioned, classic design, \\rustic, vintage, minimal technology} & \makecell{sedan, SUV, truck, convertible ; \\red, blue, black, white, silver}\\
\hline
\end{tabular}}
\captionof{table}{\textbf{Detailed Prompts for Training Sliders.} }
\label{table:detail-prompts}
\end{center}
}]

\begin{figure*}[t]
    \centering
    \includegraphics[width=1.0\linewidth]{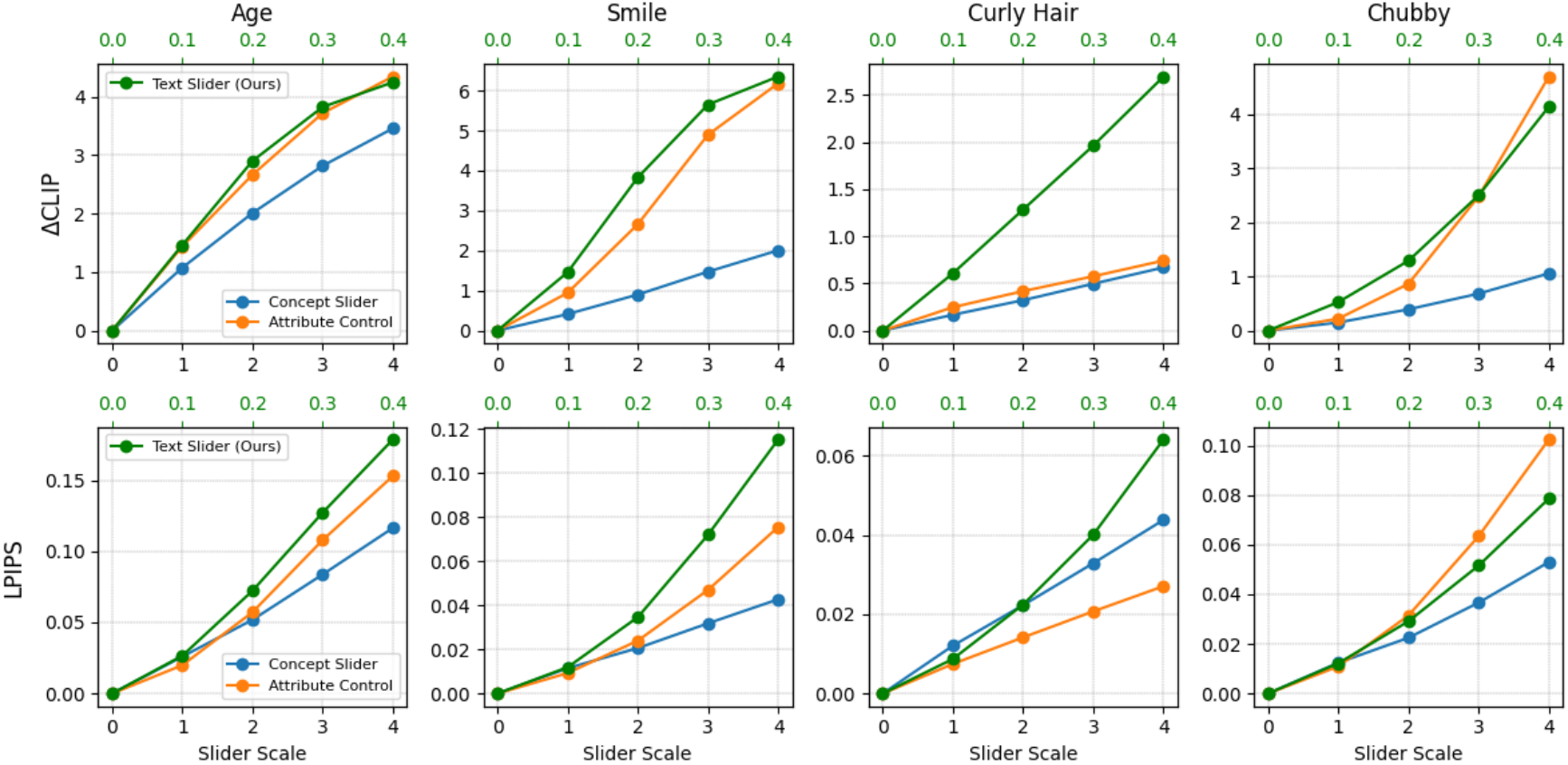}
    \caption{
\textbf{Detailed Performance of Text-to-Image Generation on SD-XL.} We report performance metrics using $\Delta$CLIP and LPIPS across four attributes—\textit{age}, \textit{smile}, \textit{curly hair}, and \textit{chubby}—evaluated at five levels of attribute intensity (slider scales). For Concept Slider~\cite{gandikota2024concept} and Attribute Control~\cite{baumann2025attributecontrol}, we assess scales from 0 to 4, while for Text Slider, we use a range of 0 to 0.4 due to its more compact scaling. Our method achieves comparable performance to the baselines while significantly reducing computational costs.}
    \label{fig:main-table-curve-t2i}
\end{figure*}

\clearpage

\end{document}